\documentclass[11pt,a4paper,UTF8,twocolumn,nopacs,floatfix,amsmath,nofootinbib,superscriptaddress,amssymb,preprintnumbers,floatfix]{revtex4}
\usepackage{setspace}
\usepackage[utf8]{inputenc}
\usepackage{makecell}
\usepackage{float}
\usepackage{amsmath,amssymb,amsfonts,bm}
\usepackage{lipsum} 
\usepackage{graphicx}
\usepackage{multirow}
\usepackage{xcolor}
\definecolor{poscolor} {RGB} {252,188,190} 
\definecolor{negcolor} {RGB} {168,168,234} 
\usepackage{slashed}
\usepackage[normalem]{ulem}
\usepackage{braket}
\usepackage{booktabs} 
\usepackage{tabularx}

\usepackage{listings} 
\usepackage{amsmath}
\usepackage{txfonts}
\usepackage{amssymb}
\usepackage{slashed}
\usepackage{color}
\usepackage{multirow}
\usepackage{float}
\usepackage{graphicx,color,dcolumn}
\usepackage{amssymb}
\usepackage{subcaption}
\usepackage{graphicx}   
\usepackage{orcidlink}
\usepackage{hyperref}
\hypersetup{colorlinks,citecolor=blue,anchorcolor=red,menucolor=red, linkcolor=red,filecolor=red,runcolor=red,urlcolor=blue,frenchlinks=true}
\usepackage{appendix}
\allowdisplaybreaks

\newcommand{\xju}{\affiliation{School of  Physical  Science and Technology ,Xinjiang University , Urumqi, Xinjiang 830046 China}}

\begin{document}
\title{Possibility of the antibottom-strange molecular pentaquarks near $ B\Sigma$ and $ B^*\Sigma$ thresholds}

\author{Jian-Kang Zhao\orcidlink{0009-0002-9927-2657}}
\xju
\author{Nijiati Yalikun\orcidlink{0000-0002-3585-1863}}
\email{nijiati@xju.edu.cn}
\xju

\begin{abstract}
 The molecular states in coupled channel system of $B_s^0N- B\Lambda- B^*\Lambda- B\Sigma- B^*\Sigma$ are investigated within one-boson-exchange model that includes $S$-$D$ wave mixing and a tunable short-range $\delta(r)$ term. Bound, resonant, and virtual states are searched by analytically continuing the $S$ matrix in the complex energy plane. In the single-channel analysis, the $ B\Sigma$ and $ B^*\Sigma$ systems with $J^P=1/2^-$ and $3/2^-$ are found to be attractive and form three bound states with reasonable cutoff region, corresponding to $1/2^-( B\Sigma)$, $1/2^-( B^*\Sigma)$, and $3/2^-( B^*\Sigma)$ . When coupled-channel dynamics is included, these states evolve into near-threshold poles that produce enhancements in the $B_s^0N$, $ B\Lambda$, and $ B^*\Lambda$ invariant mass spectra. The pole below the $ B^*\Sigma$ threshold in the $J^P=1/2^-$ system depends strongly on the treatment of the short-range $\delta(r)$ term and becomes a virtual state when this contribution is removed, while the other two resonant poles and their line shapes are only moderately affected. The influences of $B^{(*)}\Sigma^*$ channels on these states are also investigated. For representative parameter region, the predicted masses of these states lie in the $6.44-6.70$ GeV region and have widths of a few to several MeV. The pole coupling and partial-width analyses indicate that the invariant mass spectrum of the $B_s^0N$ and $ B^{(*)}\Lambda$ channels are the best place to hunt the predicted states. These results support the existence of near-threshold molecular pentaquark $P_{  \bar b s}$ and provide useful guidance in future experimental searches.
\end{abstract}
\maketitle

\section{Introduction}
\label{sec:intro}

Strong interaction described by the quantum chromodynamics (QCD) has achieved remarkable success as a fundamental component of the Standard Model of particle physics. However, its low-energy dynamics remains not fully understood, and understanding the nonperturbative behavior of QCD is essential for clarifying whether exotic hadronic configurations exist beyond the conventional quark-model classification of $qqq$ baryons and $q\bar{q}$ mesons~\cite{GellMann:1964nj,Zweig:1964jf}. Since the observation of the charmonium-like state $X(3872)$ by the Belle Collaboration~\cite{Belle:2003nnu}, many near-threshold structures have been reported in the heavy-flavor sector, stimulating extensive theoretical and experimental works to understand the role of multiquark configurations, hadronic molecules, threshold cusps, and coupled-channel dynamics; for recent reviews, see Refs.~\cite{Lebed:2016hpi,Esposito:2016noz,Chen:2016qju,Guo:2017jvc,Liu:2019zoy,Chen:2022asf,Wang:2025sic}. In particular, the observation of the hidden-charm pentaquark candidates $P_c(4380)$ and $P_c(4450)$ by the LHCb Collaboration~\cite{LHCb:2015yax}, followed by the refined structures $P_c(4312)$, $P_c(4440)$, and $P_c(4457)$~\cite{LHCb:2019kea}, as well as the hidden-charm strange states $P_{cs}(4459)$ and $P_{cs}(4338)$~\cite{LHCb:2020jpq,LHCb:2022ogu}, provides strong evidence that baryon-meson degrees of freedom play an essential role in the formation of exotic hadrons. A remarkable feature of these states is that their masses very close to relevant hadron-hadron thresholds, which naturally suggests that at least part of the observed spectrum may originate from molecular dynamics~\cite{Dong:2020hxe,Mutuk:2024ach,Brambilla:2019esw,Yamaguchi:2019vea,Dong:2021juy,Dong:2021bvy,Mai:2022eur,Zou:2021sha}.

The molecular interpretation is particularly compelling in heavy-hadron systems. Because the large mass of the heavy quark suppresses the kinetic energy, a moderate attractive interaction may be sufficient to form a shallow bound state or a near-threshold resonance. Heavy-quark spin symmetry further implies that molecular states may appear in spin-related multiplets. This mechanism has been widely used to interpret the observed $P_c$ and $P_{cs}$ states and can be naturally extended from the charm sector to the bottom and antibottom sectors~\cite{Yu:2018yxl,Huang:2018bed,Song:2024yli,Shen:2022rpn,Wang:2026prg,Suntharawirat:2026fzn}. Compared with charmed systems, bottom systems possess larger reduced masses and are generally more favorable for binding. Therefore, the investigation of bottom molecular pentaquarks provides an important opportunity to test whether the near-threshold molecular mechanism is a universal feature of heavy-flavor hadron spectroscopy.

Many studies have explored possible molecular pictures of bottom pentaquarks. The $\bar B^{(*)} N$ and $\bar B^{(*)} \Delta$ systems are investigated in Refs.~\cite{Liang:2014eba,Jian:2022rln} to understand the nature of the $\Lambda_b(6146)$ and $\Lambda_b(6152)$, and several bottom pentaquarks mass are predicted. In Ref.~\cite{Song:2025pns}, their SU(3) partners were also studied by employing the local hidden gauge approach in combination with the coupled-channel Bethe--Salpeter equation. Nevertheless, systematic studies based on meson-exchange dynamics, especially those including a detailed discussion of short-range interactions in this sector, are still needed.  Among possible bottom-flavored molecular configurations, the antibottom-strange pentaquark candidates denoted here as $P_{  \bar b s}$ are particularly interesting. The coupled $B_s^0N$, $ B^{(*)}\Lambda$, and $ B^{(*)}\Sigma$ channels share the same quark configuration and can coupled through light-meson exchange. In this system, the $ B\Sigma$ and $ B^*\Sigma$ thresholds are expected to be especially important because they are higher-lying channels that may provide strong attraction and generate poles close to threshold, in close analogy with the $\bar D\Sigma_c$ and $\bar D^*\Sigma_c$ channels associated with LHCb $P_c$ pentaquarks\cite{LHCb:2015yax,LHCb:2019kea}. At the same time, lower channels such as $B_s^0N$, $ B\Lambda$, and $ B^*\Lambda$ provide open decay modes and can slightly modify the pole positions and widths. Therefore, a reliable description of the $P_{  \bar b s}$ spectrum requires the coupled-channel calculation.

The light-meson-exchange dynamics of hadron-hadron interactions is one of the main mechanisms used to describe hadronic molecules. In the one-boson-exchange (OBE) model, the interaction between two hadrons is generated by the exchange of the SU(3) vector-nonet mesons, pseudoscalar-octet mesons, and the $\sigma$ meson. The OBE model has been widely and successfully applied to hadronic molecular candidates in the heavy-flavor sector, especially hidden-charm pentaquarks~\cite{He:2019rva,Chen:2019asm,Liu:2019zvb,Du:2021fmf,Yalikun:2021bfm,Yalikun:2023waw,Wang:2025ioe}. The effective Lagrangians used in this framework are constrained by heavy-quark spin symmetry for the heavy-meson sector and by SU(3) flavor symmetry for the light-baryon and light-meson sectors~\cite{Yan:1992gz,Wise:1992hn,Cho:1994vg,Casalbuoni:1996pg,Doring:2010ap,deSwart:1963pdg}. Therefore, the OBE model provides a useful phenomenological tool for identifying molecular states. An important issue in the OBE model is the treatment of short-range interactions. In coordinate space, the OBE potential may contain a $\delta(\bm r)$ term, which represents a contact contribution arising from the short-distance part of the interaction. In the literature, different prescriptions have been adopted: in some studies the $\delta(\bm r)$ term is retained~\cite{Liu:2009qhy,Wang:2020dya,Chen:2021tip,Wang:2022mxy,Wang:2024ukc}, while in others it is removed or absorbed into the regularized potential~\cite{Thomas:2008ja,Liu:2019zvb,Ling:2021asz,Xu:2025mhc}. Since near-threshold states are highly sensitive to short-distance dynamics, the treatment of this term may have a significant impact on the formation of molecular poles. Following the strategy introduced in our previous work~\cite{Yalikun:2021bfm}, we use a dimensionless parameter $a$ to control the strength of the $\delta(\bm r)$ contribution. This parameter effectively accounts for residual short-range dynamics, including possible effects from heavier-meson exchanges. By varying $a$, one can examine whether the predicted states are stable molecular structures or artifacts of a particular short-distance prescription.

In this work, we investigate the $B_s^0N- B\Lambda- B^*\Lambda- B\Sigma- B^*\Sigma$ system within a coupled-channel OBE model. We derive the effective interactions among heavy antimesons, light octet baryons, and exchanged light mesons using effective Lagrangians that respect heavy-quark spin symmetry(HQSS) and SU(3) flavor symmetry. A monopole form factor with cutoff $\Lambda$ is introduced to regularize the potential, while the parameter $a$ is used to adjust the contribution of the short-range $\delta(\bm r)$ term. By solving the stationary Schr\"odinger equation with $S$-$D$ wave mixing and analytically continuing the coupled-channel $S$ matrix to the complex energy plane, we search for near-threshold poles associated with the $ B\Sigma$ and $ B^*\Sigma$ channels. By varying both the cutoff $\Lambda$ and the short-range parameter $a$, this work may provides a quantitative test of the sensitivity of the predicted states to unresolved short-range dynamics. These states provide useful theoretical guidance for future experimental searches for antibottom-strange molecular pentaquarks in the $B_s^0N$ and $B^{(*)} \Lambda$ invariant mass spectra.

The rest of this paper is organized as follows. In Sec.~\ref{model}, we introduce the effective Lagrangians and derive the OBE potentials for the coupled-channel system. In Sec.~\ref{sec:numerical_results}, we present the single-channel bound-state analysis and discuss the coupled channel dynamics, including pole coupling, partial decays, and possible experimental signatures. Finally, a summary is given in the last Section.

\section{Theoretical frame}\label{model}
\subsection{The effective Lagrangian}
The OBE model has been quiet successful in describing hadronic interactions and the formation of molecular pentaquarks \cite{Du:2021fmf, He:2019rva, Chen:2019asm, Yalikun:2021bfm}. In this study, we employ it to investigate the coupled-channel dynamics of the $B_s^0N- B\Lambda- B^*\Lambda- B\Sigma- B^*\Sigma$ system and to calculate possible hadronic molecular states.

To describe the interaction between anticharmed mesons and light scalar, pseudoscalar, and vector mesons, we employ the effective Lagrangian respects HQSS and SU(3) flavor symmetry \cite{Yan:1992gz, Wise:1992hn, Cho:1994vg, Cheng:1992xi, Pirjol:1997nh, Liu:2011xc}. The relevant vertices are as follows, 
\begin{align}
	\mathcal{L}_{\tilde{\mathcal{P}}\tilde{\mathcal{P}}\sigma}&=2g_S\tilde{\mathcal{P}}^{*\mu\dagger}_{a}\sigma \tilde{\mathcal{P}}^*_{a\mu}-2g_S\tilde{\mathcal{P}}^\dagger_a \sigma \tilde{\mathcal{P}}_a,\label{lag:ppsigma}\\
	\mathcal{L}_{\tilde{\mathcal{P}}\tilde{\mathcal{P}} V}&=-\sqrt{2}\beta g_V\tilde{\mathcal{P}}^{*\dagger}_{a\mu}v_\alpha \mathbb{V}^\alpha_{ab}\tilde{\mathcal{P}}^{*\mu}_b\notag\\
    &-i2\sqrt{2}\lambda g_V \tilde{\mathcal{P}}^{*\dagger}_{a\mu}F^{\mu\nu}_{ab}\tilde{\mathcal{P}}^*_{b\nu}+\sqrt{2}\beta g_V\tilde{\mathcal{P}}^\dagger_a v_\alpha\mathbb{V}^\alpha_{ab} \tilde{\mathcal{P}}_b\notag\\
	&+(2\sqrt{2}\lambda g_V \varepsilon^{\alpha\beta\mu\kappa}v_\kappa\tilde{\mathcal{P}}^{*\dagger}_{a\mu}\partial_\alpha\mathbb{V}_{ab\beta}\tilde{\mathcal{P}}_b+H.c.),\label{lag:ppV}\\
	\mathcal{L}_{\tilde{\mathcal{P}}\tilde{\mathcal{P}}\mathbb{P}}&=i\frac{2g}{f_\pi}\varepsilon^{\alpha\mu\nu\kappa} v_\kappa \tilde{\mathcal{P}}^{*\dagger}_{a\mu}\partial_\alpha\mathbb{P}_{ab}\tilde{\mathcal{P}}^*_{b\nu}\notag\\
    &+\frac{2g}{f_\pi}(\tilde{\mathcal{P}}^{*\dagger}_{a\mu}\partial^\mu\mathbb{P}_{ab}\tilde{\mathcal{P}}_b+H.c.),\label{lag:ppP}
\end{align}
Here $\sigma$ is the light scalar field, $a,b$ are flavor indices, and $F^{\mu\nu}=\partial^\mu \mathbb{V}^\nu-\partial^\nu \mathbb{V}^\mu+i\frac{g_V}{\sqrt{2}}[\mathbb{V}^\mu,\mathbb{V}^\nu]$ is the vector-meson field strength tensor with $[\mathbb{V}^\mu,\mathbb{V}^\nu]_-=\mathbb{V}^\mu\mathbb{V}^\nu-\mathbb{V}^\nu\mathbb{V}^\mu$. The light pseudoscalar octet and the vector nonet are denoted by $\mathbb{P}$ and $\mathbb{V}^\alpha$, which in  matrix form are shown as
	\begin{align}
		\mathbb{P}&=
		\begin{pmatrix}
			\frac{\pi^0}{\sqrt 2}+\frac{\eta}{\sqrt 6}&\pi^+&K^+\\
			\pi^-&-\frac{\pi^0}{\sqrt 2}+\frac{\eta}{\sqrt 6}&K^0\\
			K^-&\bar K^0&-\sqrt{\frac{2}{3}\eta}
		\end{pmatrix},\\
		\mathbb{V}&=
	\begin{pmatrix}
		\frac{\rho^0}{\sqrt 2}+\frac{\omega}{\sqrt 2}&\rho^+&K^{*+}\\
		\rho^-&-\frac{\rho^0}{\sqrt 2}+\frac{\omega}{\sqrt 2}&K^{*0}\\
		K^{*-}&\bar K^{*0}&\phi,
	\end{pmatrix}
	\end{align}
The scalar coupling $g_S=g_\pi/(2\sqrt{6})$ with $g_\pi=3.73$ \cite{Bardeen:2003kt, Yalikun:2021dpk}. In the present OBE framework, the \(\sigma\) field is treated as an effective SU(3)-singlet field. In the conventional \(q\bar q\) flavor basis, its singlet flavor structure is written as
$\sigma=(u\bar u+d\bar d+s\bar s)/\sqrt{3}$. 
Accordingly, the scalar interaction in Eq.~\eqref{lag:ppsigma} is flavor diagonal and contains a \(B_s^0B_s^0\sigma\) vertex through the strange component of the singlet scalar field. We stress that this singlet decomposition defines the effective \(\sigma\) field adopted in the present OBE model. The physical \(f_0(500)\) has a more complicated microscopic structure and is not established to be a pure SU(3)-singlet \(q\bar q\) state. Therefore, the scalar coupling employed here should be understood as a phenomenological parametrization of the scalar-isoscalar interaction.
The scaled anti-bottomed meson field $\tilde{\mathcal{P}}^{(*)}$ is defined in flavor/isospin space as $(B^+,B^0,B_s^0)$ and $(B^{*+},B^{0*},B_s^{0*})$ \cite{Harrison:2023dzh, LHCb:2024wnx, Machleidt:1987hj}. The pion decay constant is $f_\pi=132$~MeV, and the vector couplings are $g_V=5.9$, $\beta=0.9$, $\lambda=0.56$~GeV$^{-1}$ \cite{Bando:1987br}.

Meanwhile,the SU(3) singlet terms of the $8\otimes 8\otimes 8$ interacting vertices are embedded into the effective Lagrangian as~\cite{Pich:1995bw,Bernard:1995dp},
\begin{widetext}
\begin{align}
	\mathcal{L}_{\mathcal{B} \mathcal{B} \sigma}&=-g_{\mathcal{B}\mathcal{B}\sigma} \langle \bar{\mathcal{B}} \mathcal{B} \rangle \phi_\sigma \label{lag:BBsigma}\\
	\mathcal{L}_{\mathcal{B} \mathcal{B} V}&=-\sqrt 2D'\langle\bar{\mathcal{B}} \gamma_\mu [\tilde{\mathbb{V}}^\mu_1,\mathcal{B}]_+\rangle-\sqrt 2F'\langle\bar{\mathcal{B}}\gamma_\mu [\tilde{\mathbb{V}}^\mu_1,\mathcal{B}]_-\rangle+\frac{\sqrt 2D''}{2m_{\mathcal{B}}}\langle\bar{\mathcal{B}} \sigma_{\mu\nu}\partial^\nu [\tilde{\mathbb{V}}^\mu_2,\mathcal{B}]_+\rangle+\frac{\sqrt 2F''}{2m_{\mathcal{B}}}\langle\bar{\mathcal{B}}\sigma_{\mu\nu}\partial^\nu[\tilde{\mathbb{V}}^\mu_2,\mathcal{B}]_-\rangle\label{lag:BBV}\\
	\mathcal{L}_{\mathcal{B} \mathcal{B} P}&=-\frac{\sqrt{2}D}{m_P} \langle \bar{\mathcal{B}} \gamma^5 \gamma_\mu [\partial^\mu \mathbb{P}, \mathcal{B}]_+ \rangle - \frac{\sqrt{2}F}{m_P} \langle \bar{\mathcal{B}} \gamma^5 \gamma_\mu [\partial^\mu \mathbb{P}, \mathcal{B}]_- \rangle \label{lag:BBP}
\end{align}
\end{widetext}
where $\langle\cdots\rangle$ denotes the trace over SU(3) matrices, $[A,B]_\pm = AB \pm BA$, and $\phi_\sigma$ is the scalar field operator. $\tilde{\mathbb{V}}_i$ is the nonet vector meson matrix, in which octet $\omega_8$ and singlet $\omega_1$ states are not mixed, $D'(D'')$ and $F'(F'')$ are the two independent coupling for vector (tensor) currents. The SU(3) matrix representations of $\tilde{\mathbb{V}}_i$ and $\mathcal{B}$ are as follows
\begin{align}
	\tilde{\mathbb{V}}_i&=
	\begin{pmatrix}
		\frac{\rho^0}{\sqrt 2}+\frac{\omega_8}{\sqrt 6}&\rho^+&K^{*+}\\
		\rho^-&-\frac{\rho^0}{\sqrt 2}+\frac{\omega_8}{\sqrt 6}&K^{*0}\\
		K^{*-}&\bar K^{*0}&-\sqrt{\frac{2}{3}}\omega_8
	\end{pmatrix},\notag\\
	&+g'_i\begin{pmatrix}
		\omega_1&0&0\\
		0&\omega_1&0\\
		0&0&\omega_1
	\end{pmatrix},\\
\mathcal{B}&=
\begin{pmatrix}
\frac{\Sigma^0}{\sqrt 2}+\frac{\Lambda}{\sqrt 6} & \Sigma^+ & p \\
\Sigma^- & -\frac{\Sigma^0}{\sqrt 2}+\frac{\Lambda}{\sqrt 6} & n \\
\Xi^- & \Xi^0 & -\sqrt{\frac{2}{3}}\Lambda
\end{pmatrix},
\end{align}
where $i=1,2$, $g'_1$ and $g'_2$ describe the couplings of the singlet vector meson $\omega_1$ via vector and tensor currents. The couplings of the physical $\omega$ and $\phi$ are obtained by assuming ideal mixing of $\omega_8$ and $\omega_1$\cite{ParticleDataGroup:2026aaa}
\begin{align}
\begin{pmatrix}
	\omega_8\\\omega_1
\end{pmatrix}=
\begin{pmatrix}
	\sqrt{\frac{1}{3}}&\sqrt{\frac{2}{3}}\\\sqrt{\frac{2}{3}}&-\sqrt{\frac{1}{3}}
\end{pmatrix}
\begin{pmatrix}
	\omega\\\phi
\end{pmatrix}.
\end{align}
Furthermore, we assume that the $\phi$ meson does not couple to the nucleon (Okubo-Zweig-Iizuka rule) to fix the singlet coupling constants to be $g'_1=(3F'-D')/(2\sqrt 3 D')$ and $g'_2=(3F''-D'')/(2\sqrt 3 D'')$.

For the coupling constants, we adopt the values extracted from experimental resources. The scalar meson couplings for the octet baryons are given in Ref.~\cite{Ronchen:2012eg} as $g_{NN\sigma}=8.465$, $g_{\Lambda\Lambda\sigma}=7.579$ and $g_{\Sigma\Sigma\sigma}=10.85$. The pseudoscalar meson coupling $D$ and $F$ as well as the vector meson couplings $D'$, $F'$, $D''$ and $F''$ have relation with nucleons, and shown in Table~\ref{tab:couplings}. The value of the scalar meson couplings as well as the nucleon couplings $g_{NN\pi}$, $g_{NN\rho}$, and $f_{NN\rho}$ are determined by the fit of $\pi N$-scattering data in Ref~\cite{Ronchen:2012eg}. The relative signs among the couplings are determined using the quark model, details are given at the Appendix B in Ref.\cite{Yalikun:2021dpk}. Specifically, $g_{NN\pi}$ and $g$ have opposite signs, while $g_{NN\rho}$ and $\beta$, $\kappa_{NN\rho}$ and $\lambda$, and $g_{NN\sigma}$ and $g_S$ have the same signs.

\begin{table}[htbp!]
\caption{Coupling constants for octet baryons with light mesons. The values are calculated from the nucleon couplings in Refs.~\cite{deSwart:1963pdg,Doi:2002qg,Adamuscin:2016rer}.}
\label{tab:couplings}
\begin{ruledtabular}
\begin{tabular}{ c c c }
 coupling  & Value         & Relation with nucleon \\
\midrule
 $D$     & $0.593$   & $D = g_{\mathrm{NN\pi}}(1-a_P)$, $a_P=0.4$ \\
 $F$     & $0.396$   & $F = g_{\mathrm{NN\pi}}\,a_P$ \\
 $D'$    & $-0.488$  & $D' = g_{\mathrm{NN}\rho}(1-a_V)$, $a_V=1.15$ \\
 $F'$    & $3.738$   & $F' = a_V\,g_{\mathrm{NN}\rho}$ \\
 $D''$   & $14.869$  & $D'' = f_{\mathrm{NN}\rho} - F''$, $f_{\mathrm{NN}\rho} = \kappa_{\mathrm{NN}\rho} g_{\mathrm{NN}\rho}$ \\
 $F''$   & $4.956$   & $F'' = \frac{1}{4} f_{\mathrm{NN}\rho}$ \\
\end{tabular}
\end{ruledtabular}
\end{table}

\subsection{The potentials}\label{sec:pot1}

The potential is obtained by matching the relativistic scattering amplitude to the nonrelativistic Born amplitude\cite{Breit:1929zz,Breit:1930zza}. For the two-body process
\begin{align}	
h_1(p_1)+h_2(p_2)\to h_3(p_3)+h_4(p_4),
\end{align}
the relativistic \(S\)-matrix is written as
\begin{align}
S_{fi}^{\rm rel}
=
\delta_{fi}
+i(2\pi)^4\delta^{(4)}(P_f-P_i)\mathcal M_{fi},
\end{align}
where \(\mathcal M_{fi}\) is the invariant amplitude calculated from the effective Lagrangians. The relativistic one-particle states are normalized as
\begin{align}
\langle \bm p'|\bm p\rangle_{\rm rel}
=
2E(2\pi)^3\delta^{(3)}(\bm p'-\bm p),
\end{align}
whereas the nonrelativistic states used in the Schrödinger equation satisfy
\begin{align}
\langle \bm p'|\bm p\rangle_{\rm NR}
=
(2\pi)^3\delta^{(3)}(\bm p'-\bm p).
\end{align}
Thus each external particle contributes a factor \(1/\sqrt{2E_i}\) when the relativistic amplitude is converted into the nonrelativistic one. In the nonrelativistic scattering theory with Born approximation, the nonrelativistic $S^{\rm{NR}}$ matrix has relation with potential $\mathcal{V}$ as 
\begin{align}
S_{fi}^{\rm NR}
=
\delta_{fi}
-i(2\pi)^4\delta^{(4)}(P_f-P_i)\mathcal V_{fi}.
\end{align}
Then, the nonrelativistic scattering amplitude is identified with the momentum-space potential, leading to
\begin{align}
\mathcal V_{fi}
=
-\frac{\mathcal M_{fi}}
{\sqrt{2E_1\,2E_2\,2E_3\,2E_4}}.
\end{align}
 Adopting the nonrelativistic limit on the external particles $(E_i\simeq M_i)$, one gets
\begin{align}
	\mathcal{V}^{h_1h_2\to h_3h_4}=-\frac{\mathcal{M}(h_1h_2 \to h_3h_4)}{\sqrt{2M_1 2M_2 2M_3 2M_4}},\label{eq:Breit}
\end{align}
where $M_i$ are the masses of the particles, and $\mathcal{M}$ is the $t$-channel scattering amplitude for the transition $h_1h_2\to h_3h_4$.
  The relation in Eq.~\eqref{eq:Breit} follows from matching the relativistic scattering amplitude to the nonrelativistic Born amplitude, with the normalization factors arising from the conversion between relativistically and nonrelativistically normalized external states. It should be emphasized that the Born approximation is employed only to construct the interaction kernel. The resulting potential is subsequently incorporated into the coupled-channel Schrödinger equation, in which coupled-channel rescattering are treated nonperturbatively. The nonrelativistic reduction is based on the expansion $E_i=M_i+\mathcal{O}(\bm p_i^2/M_i) $. Thus, replacing the external-state normalization factors by their mass-dependent counterparts introduces corrections of relative order $\bm p_i^2/M_i$. This approximation is particularly suitable for loosely bound molecular configurations near the corresponding hadron-hadron thresholds, where the relative momenta of the constituent hadrons are small compared with their masses. We retain the momentum-dependent contributions displayed in the OBE potentials derived below, but do not include a systematic treatment of all higher-order relativistic corrections. The cutoff $\Lambda$ in the form factor can characterize those phenomenological uncertainties, and its variation is regarded as a good estimate of all omitted relativistic contributions.

In the derivation of the scattering amplitude, we adopt the positive-energy Dirac spinor for the baryon $\mathcal{B}$ in the nonrelativistic approximation~\cite{Lu:2017dvm},
\begin{align}
    u(\bm p,s)=\sqrt{E+M}\begin{pmatrix}
        \chi\\
        \frac{\sigma \cdot \bm p}{E+M}\chi
    \end{pmatrix},
\end{align}
which satisfies $\bar{u}(\bm p,s)u(\bm p,s)=2M$, where $\sigma$ are the Pauli matrices and $\chi$ is a two-component spinor, for the scaled heavy meson fields $\tilde{\mathcal{P}}$ and $\tilde{\mathcal{P}}^*$, the normalization relations are~\cite{Yalikun:2021bfm,Wise:1992hn},
\begin{align}
    \langle 0 |\tilde{\mathcal{P}}| \bar c q(0^-)\rangle=\sqrt{M_{\tilde{\mathcal{P}}}},\qquad 
    \langle 0|\tilde{\mathcal{P}}^*_\mu|\bar c q(1^-)\rangle=\epsilon_\mu\sqrt{M_{\tilde{\mathcal{P}}^*}},
\end{align}
where $\epsilon_\mu$ is the polarization vector of $\tilde{\mathcal{P}}^*_\mu$. In the center-of-mass frame, the four-momenta of the particles in the initial state are $p_1=(E_1,\bm p)$ and $p_2=(E_2,-\bm p)$, while the four-momenta of the particles in the final state are $p_3=(E_3,\bm p')$ and $p_4=(E_4,-\bm p')$. The four-momentum of the exchanged meson is given by $q=p_3-p_1=p_2-p_4=(q^0,\bm q)$. For the convenience of the calculation, we define the new variables  
\begin{align}
	\bm q=\bm p'-\bm p,\ \bm Q=\frac{1}{2}(\bm p' +\bm p).
\end{align} 

The potentials of the $B_s^0N- B^{(*)}\Lambda- B^{(*)}\Sigma$ coupled channel system can be obtained from the potentials of $\tilde{\mathcal{P}}\mathcal{B}\to \tilde{\mathcal{P}}\mathcal{B}$, $\tilde{\mathcal{P}}\mathcal{B}\to \tilde{\mathcal{P}}^{*}\mathcal{B}$, and $\tilde{\mathcal{P}}^{*}\mathcal{B}\to \tilde{\mathcal{P}}^{*}\mathcal{B}$ by multiplying the isospin factor for each light meson exchange. For convenience, the SU(3) flavor-related factors in Lagrangian Eqs.~\eqref{lag:BBP} and \eqref{lag:BBV} are absorbed into the constants $x_{1-6}$ as  
\begin{align}
	\mathcal{L}_{\mathcal{B} \mathcal{B} P}&=-\frac{g_1}{m_P}  \bar{\mathcal{B}} \gamma^5 \gamma_\mu \partial^\mu \mathbb{P} \mathcal{B}, \label{eq:BBp}\\
	\mathcal{L}_{\mathcal{B} \mathcal{B} V}&=-g_2\bar{\mathcal{B}} \gamma_\mu {\mathbb{V}}^\mu\mathcal{B}+\frac{g_3}{2m_{\mathcal{B}}}\bar{\mathcal{B}} \sigma_{\mu\nu}\partial^\nu {\mathbb{V}}^\mu\mathcal{B},\label{eq:BBv}
\end{align}
where $g_1=\sqrt{2}(Dx_1+Fx_2)$, $g_2=\sqrt 2(D'x_3+F'x_4)$, and $g_3=\sqrt 2(D''x_5+F''x_6)$.
Than the potentials for these three scattering processes are
\begin{widetext}
\begin{subequations}\label{eq:poten-in-p}
	\begin{itemize}
		\item  $\tilde{\mathcal{P}}\mathcal{B}\to \tilde{\mathcal{P}}\mathcal{B}$
		\begin{align}
			\mathcal{V}^\sigma_1(\bm q,\bm Q) &= -\tau_\sigma g_{\mathcal{B}\mathcal{B}\sigma}g_S [ 1-\frac{i\bm\sigma\cdot(\bm q\times\bm Q)}{4 m_{\mathcal{B}}^2}] \frac{1}{\bm q^2+\mu_\sigma^2},\\
			\mathcal{V}^{\mathbb{V}}_1(\bm q,\bm Q) &=-\tau_V\frac{\beta g_V}{2} [g_2+\frac{g_2+2g_3}{4m_{\mathcal{B}}^2}i\bm\sigma\cdot(\bm q \times \bm Q)] \frac{1}{\bm q^2+\mu_V^2},
			\end{align}
		\item  $\tilde{\mathcal{P}}\mathcal{B}\to \tilde{\mathcal{P}^*}\mathcal{B}$
		\begin{align}
			\mathcal{V}^{\mathbb{P}}_2(\bm q,\bm Q) &= \tau_P\frac{gg_1}{\sqrt 2 f_\pi m_P} \frac{\bm\sigma\cdot \bm q \bm \epsilon_4^*\cdot \bm q}{\bm q^2+\mu_{P}^2} ,\\
			\mathcal{V}^{\mathbb{V}}_2(\bm q,\bm Q) &=\tau_V\frac{\lambda g_V}{2m_{\mathcal{B}}}  [(2g_2+3g_3)i\bm\epsilon_4^*\cdot (\bm q\times \bm Q) +(g_2+g_3)(\bm q^2\bm\epsilon_4^*\cdot \bm\sigma -\bm\sigma\cdot \bm q \bm\epsilon_4^*\cdot \bm q) ] \frac{1}{\bm q^2+\mu_V^2},
			\end{align}
		\item $\tilde{\mathcal{P}}^*\mathcal{B}\to \tilde{\mathcal{P}}^*\mathcal{B}$
		\begin{align}
			\mathcal{V}^\sigma_3(\bm q,\bm Q) &= -\tau_\sigma g_{\mathcal{B}\mathcal{B}\sigma}g_S\bm{\epsilon}^*_4 \cdot \bm{\epsilon}_2 [ 1-\frac{i\bm\sigma\cdot(\bm q\times\bm Q)}{4 m_{\mathcal{B}}^2} ] \frac{1}{\bm q^2+\mu_\sigma^2},\\
			\mathcal{V}^{\mathbb{P}}_3(\bm q,\bm Q) &= \tau_P\frac{gg_1}{\sqrt 2 f_\pi m_P} \frac{\bm\sigma\cdot \bm q \mathcal{T}\cdot \bm q}{\bm q^2+\mu_{P}^2} ,\\
			\mathcal{V}^{\mathbb{V}}_3(\bm q,\bm Q) &=-\tau_V \frac{\beta g_V}{ 2} [ g_2+\frac{g_3}{2m_{\mathcal{B}}^2}i\bm\sigma\cdot(\bm q \times \bm Q)] \frac{\bm{\epsilon}^*_4 \cdot \bm{\epsilon}_2}{\bm q^2+\mu_V^2}\nonumber\\
			& -\tau_V\frac{\lambda g_V }{ 2m_{\mathcal{B}}}  [(2g_2-g_3)i\mathcal{T}\cdot (\bm q\times \bm Q) -(g_2+g_3)(\bm\sigma\times\bm q)\cdot(\mathcal{T}\times \bm q) ] \frac{1}{\bm q^2+\mu_V^2}\label{eq:VIII-in-p},
			\end{align}
	\end{itemize}
	\end{subequations}
\end{widetext}
where $\tau_{\rm{ex}}$ is the isospin factor for exchanged meson, $\mu_{\rm{ex}}$ represents the effective mass of the exchanged meson, $\mu_{\rm{ex}}=\sqrt{m^2_{\rm{ex}}-(q^0)^2}$ with the energy of the exchanged meson $q^0$, $\mathcal{T}$ is defined as $\mathcal{T}=i\bm\epsilon_2\times\bm\epsilon_4^*$. With the potentials in Eq~\eqref{eq:poten-in-p}, the potentials for the $B_s^0N- B^{(*)}\Lambda- B^{(*)}\Sigma$ system can be explicitly shown in Table~\ref{tab:poten-in-p}.

\begin{table*}[htbp!]\centering
	\caption{OBE potentials of $B_s^0N- B^{(*)}\Lambda- B^{(*)}\Sigma$ coupled channel system with $I=1/2$ and the values of $x_{1-6}$ coefficients in Eqs.~\eqref{eq:BBp} and ~\eqref{eq:BBv}.}\label{tab:poten-in-p}  
	\begin{ruledtabular}
	\begin{tabular}{l c c  c c c c c c }
		Transition &  $\sum\limits_{\rm{ex}}\tau_{\rm{ex}}\mathcal{V}^{\rm{ex}}_i$ & $x_1$ & $x_2$ & $x_3$ & $x_4$ & $x_5$ & $x_6$ & $g_{\mathcal{B}\mathcal{B}\sigma}$  \\
		$ B^*\Sigma\to B^*\Sigma$ & $\mathcal{V}_3^{\sigma}-2\mathcal{V}_3^\rho+\mathcal{V}_3^\omega-2\mathcal{V}_3^\pi+\frac{1}{2\sqrt3}\mathcal{V}_3^\eta$ & 0 & $\sqrt2$& 0 & $\sqrt2$&$\frac{1}{2\sqrt2}$ & $\frac{1}{2\sqrt2}$&$g_{\Sigma\Sigma\sigma}$\\
		$ B^*\Lambda\to B^*\Sigma$ & $\sqrt3\mathcal{V}_3^{\rho}-\sqrt3\mathcal{V}_3^\pi$ & $\sqrt{\frac{2}{3}}$ & $0$ & $\sqrt{\frac{2}{3}}$ & $0$ & $\frac{\sqrt6}{4}$ & $\frac{\sqrt6}{4}$ & $0$\\
		$ B\Sigma\to B^*\Sigma$ & $-2\mathcal{V}_2^\rho+\mathcal{V}_2^\omega-2\mathcal{V}_2^\pi+\frac{1}{\sqrt3}\mathcal{V}_2^\eta$ & $\frac{\sqrt2}{3}$ & $-2\sqrt2$ & $0$ & $-\sqrt2$ & $-\frac{\sqrt2}{4}$ & $-\frac{\sqrt2}{4}$ & $0$\\
		$ B\Lambda\to B^*\Sigma$ & $\sqrt3\mathcal{V}_2^{\rho}-\sqrt3\mathcal{V}_2^\pi$ & $\sqrt{\frac{2}{3}}$ & $0$ & $\sqrt{\frac{2}{3}}$ & $0$ & $\frac{1}{2\sqrt6}$ & $\frac{1}{2\sqrt6}$ & $0$\\
		$B_s^0N\to B^*\Sigma$ & $\sqrt6\mathcal{V}_2^{\bar K^*}-\sqrt6\mathcal{V}_2^{\bar K}$ & $\frac{1}{\sqrt2}$ & $-\frac{1}{\sqrt2}$ & $\frac{1}{\sqrt2}$ & $-\frac{1}{\sqrt2}$ & $\frac{1}{2\sqrt2}$ & $\frac{1}{2\sqrt2}$ & $0$\\
		$ B^*\Lambda\to B^*\Lambda$ & $\mathcal{V}_3^{\sigma}+\mathcal{V}_3^\omega+\frac{1}{\sqrt3}\mathcal{V}_3^\eta$& $-\sqrt{2}$ & $0$ & $-\frac{2\sqrt2}{3}$ & $\sqrt{2}$ & $-\frac{\sqrt2}{4}$ & $-\frac{\sqrt2}{4}$ & $g_{\Lambda\Lambda\sigma}$\\
		$ B\Sigma\to B^*\Lambda$ & $\sqrt3\mathcal{V}_2^{\rho}-\sqrt3\mathcal{V}_2^\pi$ & $\sqrt{\frac{2}{3}}$ & $0$ & $\sqrt{\frac{2}{3}}$ & $0$ & $\frac{\sqrt6}{4}$ & $\frac{\sqrt6}{4}$ & $0$\\
		$ B\Lambda\to B^*\Lambda$ & $\sqrt3\mathcal{V}_2^{\omega}-\frac{1}{\sqrt3}\mathcal{V}_2^\eta$ & $-\sqrt{\frac{2}{3}}$ & $0$ & $-\frac{2\sqrt2}{3}$ & $\frac{5\sqrt2}{3}$ & $-\frac{\sqrt2}{4}$ & $-\frac{\sqrt2}{4}$ & $0$\\
		$B_s^0N\to B^*\Lambda$ & $\sqrt2\mathcal{V}_2^{\bar K^*}-\sqrt2\mathcal{V}_2^{\bar K}$ & $-\frac{\sqrt6}{6}$ & $-\frac{\sqrt6}{2}$ & $-\frac{\sqrt6}{6}$ & $-\frac{\sqrt6}{2}$ & $-\frac{\sqrt6}{4}$ & $-\frac{\sqrt6}{4}$ & $0$\\
		$ B\Sigma\to B\Sigma$ & $\mathcal{V}_1^{\sigma}-2\mathcal{V}_1^\rho+\mathcal{V}_1^\omega$& $0$ & $0$ & $0$ & $-\sqrt{2}$ & $-\frac{1}{2\sqrt2}$ & $-\frac{1}{2\sqrt2}$ & $g_{\Sigma\Sigma\sigma}$\\
		$ B\Lambda\to B\Sigma$ & $\sqrt3\mathcal{V}_1^\rho$& $0$ & $0$ & $\frac{\sqrt2}{3}$ & $0$ & $\frac{1}{2\sqrt2}$ & $\frac{1}{2\sqrt2}$ & $0$\\
		$ B_s^0N\to B\Sigma$ & $\sqrt6\mathcal{V}_1^{\bar K^*}$ & $0$&$0$ & $\sqrt{3}$ & $-\sqrt{3}$ & $\frac{1}{2\sqrt2}$& $\frac{1}{2\sqrt2}$& $0$\\
		$ B\Lambda\to B\Lambda$ & $\mathcal{V}_1^{\sigma}+\mathcal{V}_1^\omega$ & $0$ & $0$ & $-\frac{2\sqrt2}{3}$ &$\sqrt2$ &$-\frac{\sqrt2}{4}$&$-\frac{\sqrt2}{4}$ & $g_{\Lambda\Lambda\sigma}$\\
		$ B_s^0N\to B\Lambda$ & $\sqrt2\mathcal{V}_1^{\bar K^*}$ & $0$ & $0$ & $-\frac{1}{\sqrt3}$ & $-\sqrt{3}$ & $-\frac{\sqrt6}{4}$ & $-\frac{\sqrt6}{4}$  & $0$ \\
		$ B_s^0N\to  B_s^0N$ & $\mathcal{V}_1^\sigma$ & $0$ & $0$ & $0$ & $0$ & $0$ & $0$ & $g_{NN\sigma}$\\
	\end{tabular}
\end{ruledtabular}
\end{table*}

The potential in the momentum space in Eq.~\eqref{eq:poten-in-p} is transformed into the coordinate space potential by Fourier transform \cite{Chen:2016qju, Liu:2019zoy, Wang:2020dya},
\begin{align}
	\mathcal{V}_i^{\rm{ex}}(\bm r,\Lambda,\mu_{\rm{ex}})=\int\frac{\mathrm{d}^3 \bm q}{(2\pi)^3}\mathcal{V}_i^{\rm{ex}}(\bm q,\bm Q)F^2(\bm q,\Lambda,\mu_{\rm{ex}})e^{iq\cdot r},
\end{align}
where $F^2(\bm q,\Lambda,\mu_{\rm{ex}})$ is the form factor, which reduces the off-shell effects of the exchange meson and represents the internal structure of the interaction vertex \cite{Tornqvist:1993ng}. Based on the discussion of form factors in Ref.~\cite{Chen:2017vai}, the form factors generally take  monopole, dipole and exponential forms. In the case of low energy scale, the hadronic molecule should be almost unaffected by the type of form factor. In this study, we use the form factor in the monopole form  
\begin{align}
F(\bm q,\Lambda,\mu_{\rm{ex}})=\frac{\Lambda^2-\mu^2_{\rm{ex}}}{\Lambda^2+\bm q^2}.    
\end{align}
Since the Fourier transformation is only depending on the momentum $\bm q$ and $\bm Q$, it is sufficient to consider the following Fourier transformations to obtain the coordinate space potential
\begin{widetext}
 \begin{align}
	\frac{1}{(2\pi)^3}\int \frac{1}{\bm q^2+\mu_{\rm{ex}}} \left (\frac{\Lambda^2-\mu^2_{\rm{ex}}}{\bm q^2+\Lambda^2}\right )^2 e^{i\bm q\cdot \bm r}d^3\bm q&=\frac{1}{4\pi r}({\rm{e}}^{-\mu_{\rm{ex}}r}-{\rm{e}}^{-\Lambda r})-\frac{\Lambda ^2-\mu_{\rm{ex}}^2}{8\pi\Lambda }{\rm{e}}^{-\Lambda r}=Y_{\rm{ex}},\\
	\frac{1}{(2\pi)^3}\int\frac{i \bm A\cdot (\bm q\times\bm Q)  }{\bm q^2+\mu^2_{\rm{ex}}}\left (\frac{\Lambda ^2-\mu^2_{\rm{ex}}}{\bm q^2+\Lambda ^2}\right )^2 e^{i\bm q\cdot \bm r}d^3\bm q	&=\bm A\cdot \bm L\frac{1}{r}\frac{\partial }{\partial r} Y_{\rm{ex}},\\
	\frac{1}{(2\pi)^3}\int \frac{\bm A\cdot \bm q \bm B\cdot \bm q}{\bm q^2+\mu_{\rm{ex}}^2} \left (\frac{\Lambda ^2-\mu^2_{\rm{ex}}}{\bm q^2+\Lambda ^2}\right )^2 e^{i\bm q\cdot \bm r}d^3\bm q	&=-\frac{1}{3}[\bm A\cdot \bm BC_{\rm{ex}}+S(\bm A,\bm B,\hat r)T_{\rm{ex}}],
	 \end{align} 
	 \end{widetext}
where $\bm A$ and $\bm B$ represent the spin operators as $\bm \sigma$ and $\mathcal{T}$, $\bm L$ is the angular momentum operator, $\bm L=\bm r\times \bm Q$ ~\cite{Liu:2011xc}. $S(\bm A,\bm B,\hat r)=3 \bm A\cdot \hat r \bm B\cdot \hat r-\bm A\cdot\bm B$ is the tensor operator in coordinate space. Here, we follow the results discussed in Refs.~\cite{ Yalikun:2021bfm, Yalikun:2023waw} regarding the $\delta(\bm r)$\footnote{The form factor regularizes the entire meson-exchange potential, while the $\delta(r)$ term is a short-range component of the same potential arising from the Fourier transformation; they operate at different levels and do not introduce any double counting, following the standard prescription in Ref.~\cite{Yalikun:2021bfm}} term. The role of the $\delta(\bm r)$ term in the Fourier transformation of $\bm A\cdot \bm q \bm B\cdot \bm q/(\bm q^2+m_{\rm{ex}}^2)$ can be completely controlled once a dimensionless parameter $a$ is introduced in $C_{\rm{ex}}$, thus the functions $C_{\rm{ex}}$ and $T_{\rm{ex}}$ read
\begin{align}
C_{\rm{ex}}&=\frac{1}{r^2}\frac{\partial}{\partial r}r^2\frac{\partial}{\partial r}Y_{\rm{ex}}+\frac{a}{(2\pi)^3}\int \left (\frac{{\Lambda}^2-m^2_{\rm{ex}}}{\bm q^2+{\Lambda}^2}\right )^2 e^{i\bm q\cdot \bm r}d^3\bm q,\label{eq:Cr}\\
T_{\rm{ex}}&=r\frac{\partial}{\partial r}\frac{1}{r}\frac{\partial}{\partial r}Y_{\rm{ex}}.\label{eq:Tr}
 \end{align}   
 In this way, the contribution of the $\delta(\bm r)$ term is fully included (excluded) when $a=0(1)$. Similarly, the Fourier transformation of the function $(\bm A\times \bm q) \cdot(\bm B\times \bm q)/(\bm q^2+m^2_{\rm {ex}})$ can be evaluated with the help of the relation $(\bm A\times \bm q) \cdot(\bm B\times \bm q)=\bm A\cdot \bm B\bm q^2-\bm A\cdot \bm q \bm B\cdot \bm q$. 
For the masses of exchanged mesons, we take the isospin-averaged masses as $m_\pi=137.2$, $m_\eta=547.9$, $m_\rho=775.3$, $m_\omega=782.7$, $m_{\bar{K}}=493.7$, $m_{\bar{K}^*}=891.7$~MeV \cite{ParticleDataGroup:2026aaa}.

For the \(B_s^0N\) diagonal interaction, Table~\ref{tab:poten-in-p} contains only the \(\sigma\)-exchange contribution. Under the SU(3)-singlet prescription introduced above, the strange component of the effective \(\sigma\) field permits the \(B_s^0B_s^0\sigma\) vertex, while the \(NN\sigma\) coupling provides the vertex on the nucleon side. Hence a nonvanishing \(B_s^0N\rightarrow B_s^0N\) scalar-exchange potential is generated in the present model. We note that this contribution depends on the assumed singlet treatment of the effective scalar field; for a purely nonstrange scalar field, a direct \(B_s^0B_s^0\sigma\) coupling would instead be absent or strongly suppressed. The \(\sigma\)-exchange contribution to \(B_s^0N\) should therefore be regarded as model dependent.

In this work, we focus on the negative parity states in $B_s^0N- B^{(*)}\Lambda- B^{(*)}\Sigma$ systems which are possibly bound in $S$ wave and thus more easily form the molecular states compared to positive ones. The partial waves of the channels corresponding to the spin parities for $J^P=1/2^-$ and $3/2^-$ as well as their threshold are summarized in Table~\ref{tab:mass-parwave}. The notation $^{2S+1}L_J$ is used to identify various partial waves, in which $S$, $L$ and $J$ stand for the spin, orbital and total angular momentums, respectively. In the actual calculation, the potentials should be projected out, and this is done by sandwiching them between the partial waves of the initial and final states.We refer to Refs.~\cite{Yalikun:2021bfm,Yalikun:2021dpk} to compute the partial wave projections.
\begin{table*}[htbp!]\centering
	\caption{Thresholds of the channels and partial wave components for $J^P = 1/2^-$ and $3/2^-$ states.}\label{tab:mass-parwave}
	\begin{ruledtabular}
	\begin{tabular}{l c c c c c}
		Channels & $ B_s^0N$ & $ B\Lambda$ & $ B^\ast\Lambda$ &  $ B\Sigma$& $ B^\ast\Sigma$ \\
		\midrule
		$W_j$\,[MeV]\cite{ParticleDataGroup:2026aaa} & $6305.8$ & $6395.3$ &$6440.4$&  $6472.8$  & $6517.9$ \\
		$J^P=1/2^-$ & $^2S_{1/2}$ & $^2S_{1/2}$ & $^2S_{1/2}$, $^4D_{1/2}$  & $^2S_{1/2}$& $^2S_{1/2}$, $^4D_{1/2}$ \\
		$J^P=3/2^-$ & $^2D_{3/2}$ & $^2D_{3/2}$ & $^4S_{3/2},^2D_{3/2},^4D_{3/2}$ &$^2D_{3/2}$  & $^4S_{3/2},^2D_{3/2},^4D_{3/2}$ \\
	\end{tabular}
\end{ruledtabular}
\end{table*}

\begin{figure*}[htbp!]
	\centering 
	\includegraphics[width=0.9\textwidth]{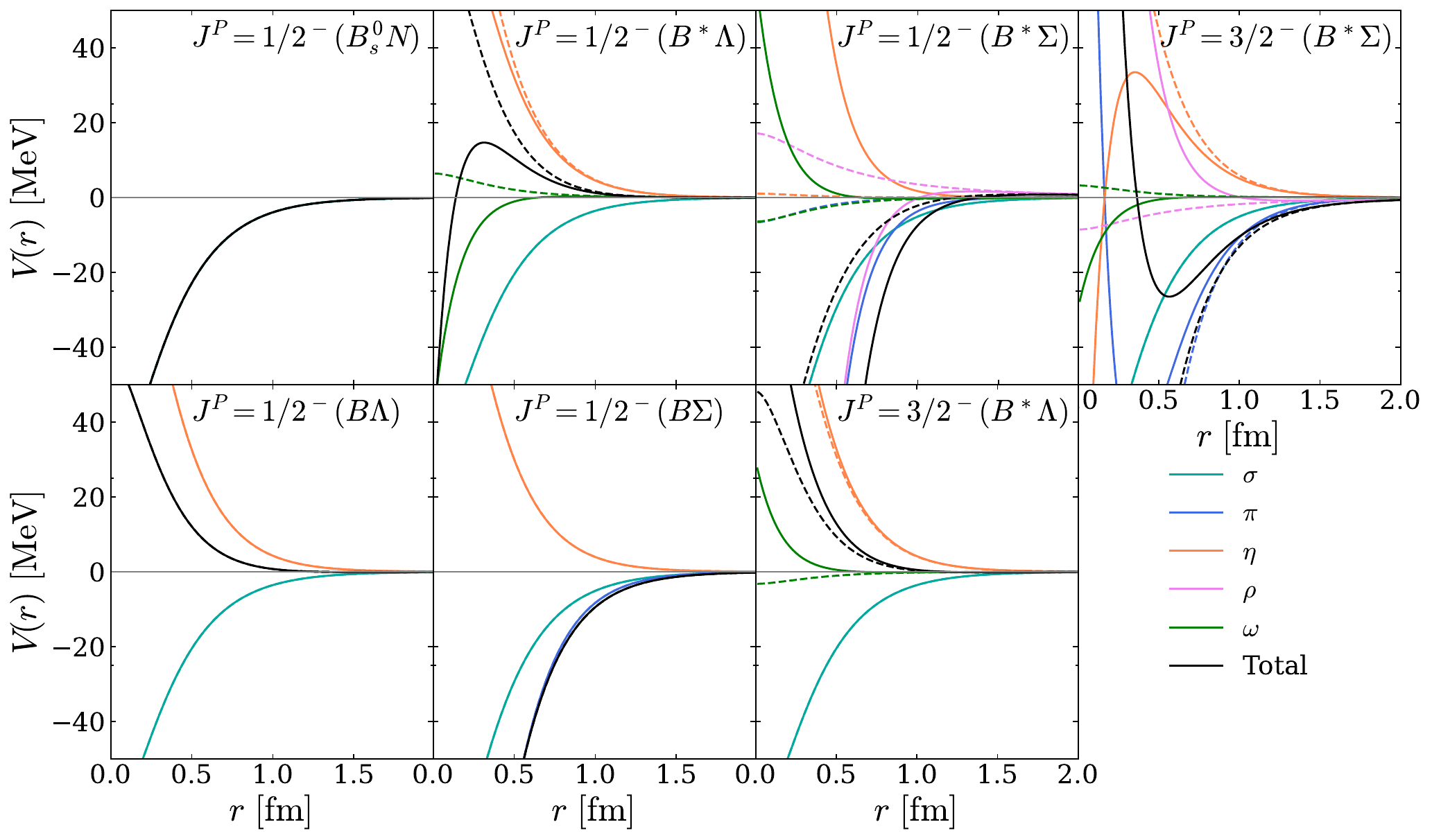} 
	\caption{Potentials of the $S$-wave states of the $ B^{(*)}\Sigma$ system with $I=1/2$ for $J^P = 1/2^-$ and $3/2^-$, with the cutoff set to $\Lambda = 1.2$~GeV. The solid and dashed curves correspond to the cases $a=0$ and $a=1$, respectively.}
	\label{V(r)_I=0.5}
\end{figure*}

Having derived the OBE potentials for each channel, we now examine the behavior of the various meson-exchange potentials under the two extreme treatments of the $\delta(\bm r)$ term. The numerical analysis centers on this behavior and depends on two sets of key parameters. Since $S$-wave interaction potentials are crucial for the formation of hadronic molecules, we extract the OBE potentials for the $S$-wave channels of the $ B^*\Sigma$ system with spin parities $J^P = 1/2^-$ and $3/2^-$. Fig.~\ref{V(r)_I=0.5} illustrates the OBE potentials for the $ B^{(*)}\Sigma$ system with $I=1/2$. In each subplot, the solid and dashed lines correspond to the cases $a=0$ and $a=1$, which represents results of fully including and removing the $\delta(\bm r)$ term, respectively.

The potentials of the $ B\Sigma$ system are only proportional to the Yukawa term $Y_{\rm{ex}}$ and independent of the $\delta(\bm r)$ term. With $\delta(\bm r)$ term, the vector and pseudoscalar meson exchange potentials of the $ B^*\Sigma$ system with $J^P=1/2^-,3/2^-$ can change their signs once due to the short-range $\delta(\bm r)$ term in their core which has an opposite sign relative to its remaining part. After removing the $\delta(\bm r)$ term, those potentials are consistent in sign in the whole range of $r$. The S wave total potentials in both of $ B\Lambda$ and $ B^*\Lambda$ are repulsive, and there is no bound state accordingly. In addition, the S-wave potential of the $ B_s^0N$ system is attractive due to the diagonal $\sigma$-exchange interaction discussed in Sec. II.B. However, within the parameter region considered here, this attraction is insufficient to generate a bound state in the single-channel calculation. The $ B_s^0N$ channel therefore acts as a lower-lying scattering and decay channel in the coupled-channel analysis.


\subsection{Schrödinger equation}
To investigate molecular states in the $ B\Sigma$ and $ B^*\Sigma$ systems, we solve the Schrödinger equation using the OBE potentials. The radial Schrödinger equation for the coupled-channel potential matrix $V_{jk}$ is given by:
\begin{eqnarray}
	\left[-\frac{1}{2\mu_j}\frac{d^2}{d r^2}+\frac{ l_j(l_j+1)}{2\mu_jr^2} + W_j\right]u_j+
	\sum_{k}
	\mathcal{V}_{jk}u_k=
	E u_j,
	\label{eq_schro_coupl}
\end{eqnarray}
Here $j$ is the channel index, $u_j(r)=rR_j(r)$ is the reduced radial wave function with orbital angular momentum $l_j$, and $\hbar=1$. The reduced mass and threshold of channel $j$ are denoted by $\mu_j$ and $W_j$, respectively. The channel momentum is defined as
\begin{align}\label{eq:ch-mom}
	q_j(E)=\sqrt{2\mu_j(E-W_j)}.
\end{align}
Solving Eq.~\eqref{eq_schro_coupl} yields the normalized wave function for channel $j$, which satisfies the incoming boundary condition \cite{Taylor:1972pty} and has the asymptotic form
\begin{eqnarray}
	u_{j}^{k}(r)\overset{r\rightarrow \infty}{\longrightarrow} \delta_{jk}h^-_{l_j}(q_j r)-S_{jk}(E)h^+_{l_j}(q_j r),\label{eq:asym-wave}
\end{eqnarray}
where $h_l^\pm(x)$ are spherical Hankel functions and $S_{jk}(E)$ is the scattering matrix. In the complex energy plane, poles of $S_{jk}(E)$ correspond to bound, virtual, and resonant states \cite{Taylor:1972pty}. For multichannel systems with thresholds $W_1<W_2<\cdots$, $S_{jk}(E)$ is analytic except for branch points at $E=W_j$ and isolated poles. To locate these poles, the $S$ matrix is analytically continued to the complex energy plane and searched on the appropriate Riemann sheet(RS). The binding energy is defined as
\begin{align}
	\mathbb{B} = E_{\rm{pole}} - W_1,
\end{align}
Since the channel momentum $q_j$ is multivalued for $E$, each channel gives two RSs: the physical sheet ($\rm{Im}[q_j]>0$) and the unphysical sheet ($\rm{Im}[q_j]<0$). Bound states appear as poles on the physical sheet with $E_{\rm{pole}}=M$. Poles on the unphysical sheet correspond to resonances when $\rm{Re}E_{\rm{pole}}$ lies above thresholds, taking the form $E_{\rm{pole}}=M-i\Gamma/2$, where $M$ is the mass and $\Gamma$ the decay width \cite{Oset:2010tof,Wang:2025ioe}. This coupled-channel formalism respects unitarity and provides a rigorous treatment of near-threshold states \cite{Yalikun:2021bfm}.

To analyze the channel contribution on the poles, we calculate the channels coupling of the poles extracted from the residual of the $T$ matrix. Following the formulation in the review article titled ``Resonance" in RPP\cite{ParticleDataGroup:2026aaa}, the relation between $S(E)$ and $T(E)$ read
\begin{align}
	S_{nk}(E)=\delta_{nk}+2i\sqrt{\rho_n}\, T_{nk}(E)\sqrt{\rho_k},
\end{align}
where $n,k$ are channel indices. In the nonrelativistic approximation, the two-body phase space factor $\rho_j$ for channel $j$ is $\rho_j = q_j(E)/(8\pi E)$. The residue at the pole $E_{\rm{pole}}$ is
\begin{align}
	R_{jk}= \lim_{E\to E_{\rm{pole}}} (E^2-E_{\rm{pole}}^2)\, T_{jk}(E)=g_j g_k,
\end{align}
with $g_j$ the coupling to channel $j$. 
The partial decay width into an open channel is \cite{Sakai:2019qph,Garzon:2012np}
\begin{align}
	\Gamma_j=\frac{q_j(M)}{8\pi M^2}|g_j|^2,\qquad M=\operatorname{Re}(E_{\mathrm{pole}}).
\end{align}

\section{Results and discussion} 
\label{sec:numerical_results}
\subsection{Single-channel analysis}\label{sec4}

Let us start with discussing possible bound states in single channels. Since the potentials for $B_s^0N$ and $B^{(*)}\Lambda$ channels are repulsive or not attractive enogh to form a bound state, our focus is moved to bound states of the $ B\Sigma$ and $ B^*\Sigma$ systems within the OBE framework. Solving the Schrödinger equation~\eqref{eq_schro_coupl} including $S$-$D$ wave mixing, we obtain bound-state energies. The parameter $a$ controls the short-range $\delta(\bm{r})$ term, acting as a phenomenological contact term for short-range hadronic dynamics \cite{Du:2021fmf}. For the isodoublet $ B\Sigma$ and $ B^*\Sigma$ systems with $J^P = 1/2^-$ and $3/2^-$, we compute the bound-state energies by varying $\Lambda$ for selected values of $a$. Table~\ref{tab:BbSig_BE_OBE} lists the bound-state energies and corresponding rms radii from single-channel calculations as the cutoff and parameter $a$ are varied, and also compares the effects of $S$-$D$ partial-wave mixing.
Four scenarios corresponding to $a = 0$, $0.5$, $0.75$, and $1$ are considered, where $a=0$ (or $a=1$) fully includes ( or removes) the $\delta(\bm{r})$ term, and $a=0.5$, $0.75$ correspond to $50\%$ and $75\%$ attenuation, respectively.

For the $ B\Sigma$ bound state with $J^P = 1/2^-$, the binding energy is independent of $a$, showing no dependence on the $\delta(\bm{r})$ term, as the OBE potential has no such contribution. In contrast, for the $ B^*\Sigma$ states with $J^P=1/2^-$ and $3/2^-$, the binding energies are significantly affected by $a$. With increasing $a$, the $J^P=1/2^-$ binding energy decreases (shallow bound state), while the $J^P=3/2^-$ binding energy increases (deep bound state). This opposite behavior arises because the $J^P=1/2^-$ potential becomes shallower with $a$, while the $J^P=3/2^-$ potential deepens as shown in Fig.~\ref{V(r)_I=0.5}. Furthermore, for $\Lambda = 1.5$ or $1.8$~GeV and $a = 0.5$ or $0.75$, the three bound states ($ B\Sigma$ $J^P=1/2^-$, $ B^*\Sigma$ $J^P=1/2^-$, and $ B^*\Sigma$ $J^P=3/2^-$) can coexist stably. When the $\delta(\bm{r})$ term is fully included ($a=0$), the binding energies exhibit a strong dependence on $\Lambda$.
 The rms provides further support for the molecular interpretation. For the $B\Sigma$ state with $J^P=1/2^-$, the rms radii is $\sim 1.79$~fm at $\Lambda=1.25$~GeV and decreases to $\sim 0.78$~fm at $\Lambda=1.8$~GeV. For the $B^*\Sigma$ states, the RMS radii exhibit a clear correlation with the binding energy: shallow bound states (binding energy $\lesssim 10$~MeV) have radii $\gtrsim 1.5$~fm, while deeper states become more compact. Notably, the $J^P=3/2^-(B^*\Sigma)$ state with $S$-$D$ mixing has an rms radii of $0.46$~fm at $\Lambda=1.80$~GeV and $a=1.0$, and the $J^P=1/2^-(B^*\Sigma)$ state with $S$-$D$ mixing reaches $0.55$~fm at $\Lambda=1.80$~GeV and $a=0.75$.

\begin{table*}[ht!]
\begin{ruledtabular}
\setlength{\tabcolsep}{2pt}
\centering
\small
\setlength{\tabcolsep}{3.5pt}
\caption{Binding energies $\mathbb{B}$  and rms radii $\sqrt{\langle r^2 \rangle}$  of bound states in the single-channel $B\Sigma$ and $B^*\Sigma$ systems with isospin $I=1/2$, as functions of the cutoff $\Lambda$  for fixed $a$. Entries marked ``$\cdots$'' indicate that the interaction is too weak to form a bound state.}
\label{tab:BbSig_BE_OBE}

\begin{tabular}{cccccccccccc}
\toprule
\multicolumn{2}{c}{} & \multicolumn{2}{c}{$B\Sigma$} & \multicolumn{4}{c}{$B^*\Sigma$ (S-wave)} & \multicolumn{4}{c}{$B^*\Sigma$ (S-D mixing)} \\
\cmidrule(lr){3-4} \cmidrule(lr){5-8} \cmidrule(lr){9-12}
& & \multicolumn{2}{c}{$J^P=1/2^-$} & \multicolumn{2}{c}{$1/2^-$} & \multicolumn{2}{c}{$3/2^-$} & \multicolumn{2}{c}{$1/2^-$} & \multicolumn{2}{c}{$3/2^-$} \\
\cmidrule(lr){3-4} \cmidrule(lr){5-6} \cmidrule(lr){7-8} \cmidrule(lr){9-10} \cmidrule(lr){11-12}
$a$ & $\Lambda$ (GeV) & $\mathbb{B}$(MeV) & $\sqrt{\langle r^2\rangle}$(fm)& $\mathbb{B}$ & $\sqrt{\langle r^2\rangle}$ & $\mathbb{B}$ & $\sqrt{\langle r^2\rangle}$ & $\mathbb{B}$ & $\sqrt{\langle r^2\rangle}$ & $\mathbb{B}$ & $\sqrt{\langle r^2\rangle}$ \\
	\midrule[0.8pt]
\multirow{4}{*}{0.0} & 1.00 & $\cdots$ & $\cdots$     & $-28.38$  &$0.81$& $\cdots$ & $\cdots$    & $-31.00$ &$0.78$ &$\cdots$  &$\cdots$ \\
                     & 1.25 & $-45.19$ & $1.79$  & $-86.13$  &$0.62$& $\cdots$ & $\cdots$    &$-94.91$  & $0.42$  & $\cdots$ &$\cdots$ \\
                     & 1.50 & $-54.78$ & $1.16$  & $-90.39$  &$0.49$& $\cdots$ & $\cdots$   & $-95.3$  &  $0.35$ & $-2.15$  & $1.64$ \\
                     & 1.80 & $-80.37$ &$0.78$   & $-125.44$ &$0.21$& $-0.15$  & $1.84$ & $-108.06$  &$0.26$   & $-8.90$  & $1.34$ \\
\midrule[0.8pt]
\multirow{4}{*}{0.5} & 1.00 & $\cdots$  &$\cdots$    & $\cdots$   & $\cdots$     & $\cdots$ & $\cdots$    & $\cdots$  & $\cdots$    & $\cdots$ & $\cdots$\\
                     & 1.25 &  $-45.19$ &$1.79$  & $-24.46$   & $0.83$  & $\cdots$ & $\cdots$   & $-29.33$  & $0.79$ & $-2.18$  & $ 1.62$ \\
                     & 1.50 & $-54.78$  &$1.16$  & $-83.45$   & $0.66$    & $-2.37$  & $1.55$ & $-84.40$   & $0.56$   & $-13.58$ &  $1.18$ \\
                     & 1.80 &$-80.37$   &$0.78$  & $-90.02$   & $0.43$    & $-11.72$ & $1.17$ & $-91.61$  & $0.38$   & $-33.51$ &   $0.90$ \\
\midrule[0.8pt]
\multirow{4}{*}{0.75}& 1.00 & $\cdots$ & $\cdots$& $\cdots$  & $\cdots$& $\cdots$ & $\cdots$ & $\cdots$  & $\cdots$ & $\cdots$ & $\cdots$ \\
                     & 1.25 & $-45.19$&$1.79$& $\cdots$  & $\cdots$ & $-0.52$  &$1.73$ & $-0.27$ & $1.78$ & $-6.50$ &$1.38$ \\
                     & 1.50 &$-54.78$ &$1.16$& $-9.64$  & $1.08$ & $-11.68$& $1.16$ & $-16.11$ & $ 0.96$& $-28.19$& $0.93$ \\
                     & 1.80 &$-80.37$ & $0.78$ & $-40.36$ & $  0.57$ & $-35.54$& $0.82$ & $-70.03$ & $0.55$& $-64.51$& $0.71$ \\
\midrule[0.8pt]
\multirow{4}{*}{1.0} & 1.00 & $\cdots$ & $\cdots$ & $\cdots$  &$\cdots$ & $\cdots$ & $\cdots$ & $\cdots$  & $\cdots$& $-0.02$ &$1.91$ \\
                     & 1.25 & $-45.19$  & $1.79$& $\cdots$  & $\cdots$ & $-5.13$  &$1.37$ & $\cdots$  & $\cdots$ & $-14.84$& $ 1.13$ \\
                     & 1.50 & $-54.78$ &$1.16$ & $\cdots$  & $\cdots$ & $-33.48$& $0.82$ & $\cdots$  & $\cdots$ & $-54.05$& $0.78$ \\
                     & 1.80 & $-80.37$ & $0.78$& $\cdots$  & $\cdots$ & $-79.35$&$0.49$ & $\cdots$  & $\cdots$ & $-92.45$& $0.46$ \\
\bottomrule
\end{tabular}
\end{ruledtabular}
\end{table*}

\subsection{Molecular states in coupled-channel systems}\label{sec5}
We solved the coupled-channel Schr\"odinger equation in Eq.~\eqref{eq_schro_coupl} for the $B_s^0N- B\Lambda- B^\ast\Lambda- B\Sigma- B^\ast\Sigma$ system, obtaining the energy-dependent $S(E)$ matrix. Its poles are searched by varying the cutoff and parameter $a$, and the pole coupling $g_i$ is calculated from the residual of the $T$ matrix. Because only the $ B^{(*)}\Sigma$ channels with $j^P=1/2^-$ and $3/2^-$ can form the bound states, we mainly discuss the poles in these systems. 

\begin{figure*}[ht!]
    \centering
    \includegraphics[width=0.9\textwidth]{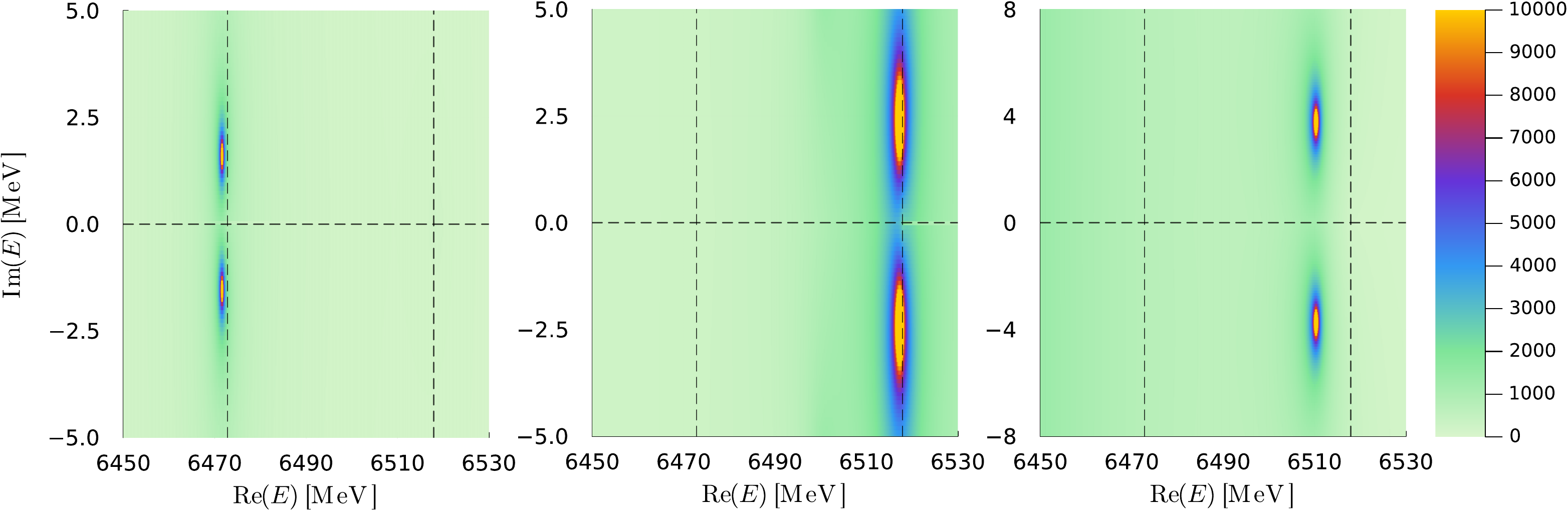}
    \caption{The $T(E)$ matrix on the complex energy plane for the full coupled-channel systems with $J^P=1/2^-$ and $3/2^-$ when $\Lambda=1.25$ GeV and $a=0$. Left and middle ones are the results for $1/2^-$ systems on $(---++)$ and $(+++--)$ RSs, while the right one is for $J^P=3/2^-$ systems on $(----+)$ RS. The vertical dashed lines mark the thresholds of $ B\Sigma$ and $ B^*\Sigma$ channels, respectively.}
    \label{fig:pole_pos}
\end{figure*}

First, we search for poles of the $T(E)$ matrix on the complex energy plane for the case $a=0$ by varying $\Lambda$. At $\Lambda=1.25$~GeV, the pole positions are shown in Fig.~\ref{fig:pole_pos}. In the $J^P=1/2^-$ system, two poles are found, located, respectively, below the $ B\Sigma$ and $ B^*\Sigma$ channel thresholds. The lower pole, situated below the $ B\Sigma$ threshold, lies on the $(---++)$ RS, which is connected to the physical real energy axis. The higher pole, located below the $ B^*\Sigma$ threshold, lies on the $(+++--)$ RS, which is remote from the physical real energy axis. In the $J^P=3/2^-$ system, one pole is found on the $(----+)$ RS that is connected to the physical real energy axis.

\begin{figure*}[ht!]
    \centering
    \includegraphics[width=0.8\textwidth]{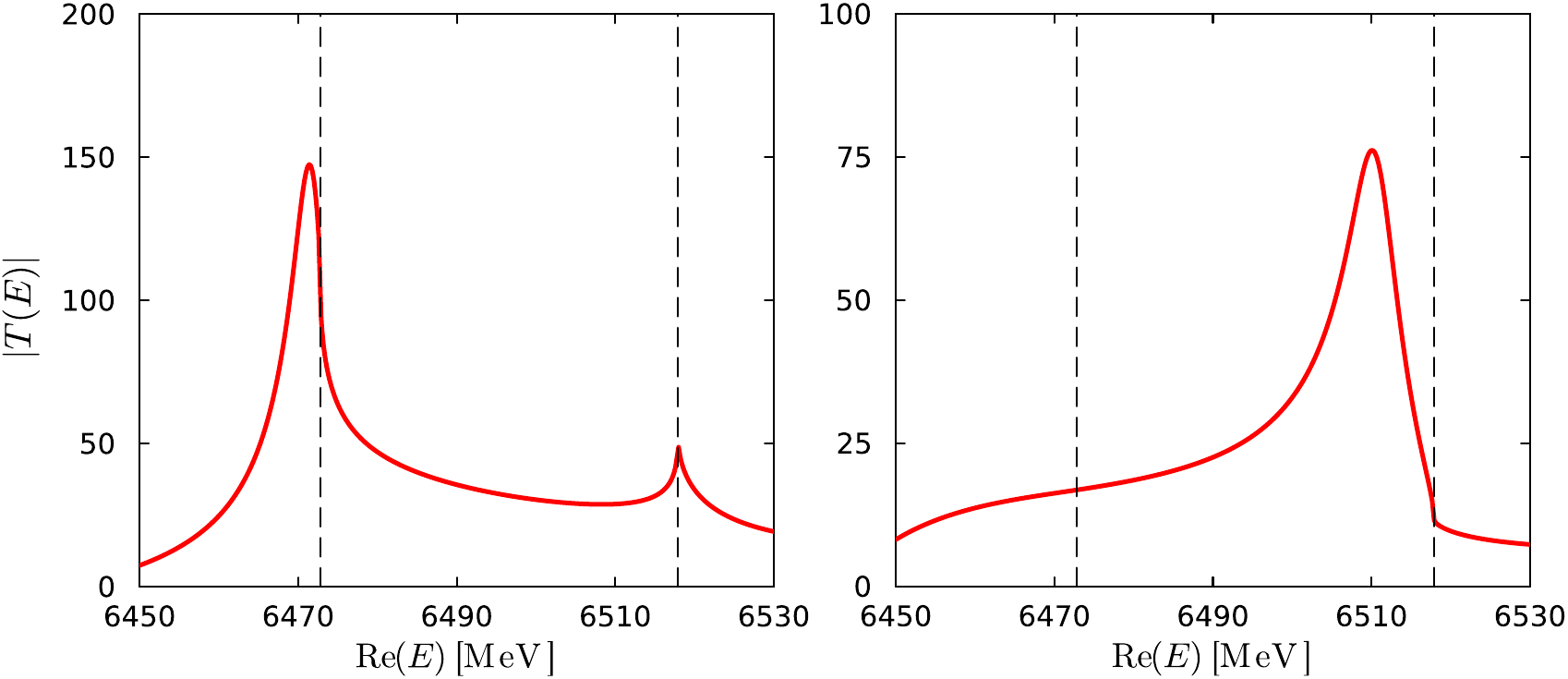}
    \caption{The absolute $T(E)$ matrix for the scattering processes $ B\Sigma(^2S_{1/2})\to B_s^0 N(^2S_{1/2})$ with $J^P=1/2^-$ (left) and $ B^*\Sigma(^4S_{3/2})\to B_s^0 N(^2D_{3/2})$ with $J^P=3/2^-$ (right), when $\Lambda=1.25$ GeV and $a=0$.}
    \label{fig:t_mat_1}
\end{figure*}

The Absolute value of $T(E)$ on the physical real energy axis for the full coupled-channel systems with $J^P=1/2^-$ and $3/2^-$ is shown in Fig.~\ref{fig:t_mat_1}. The lower pole in the $J^P=1/2^-$ system produces a peak below the $ B\Sigma$ channel threshold, while the higher pole in this system generates a cusp at the $ B^*\Sigma$ threshold because it lies on a RS not connected to the physical real energy axis. For the $J^P=3/2^-$ system, the pole creates a peak in the scattering amplitude at energies below the $ B^*\Sigma$ threshold. These features confirm that poles located on RS connected to the physical real energy axis manifest as peaks in the $T(E)$ line shape.

Now, we analyze the effects of $\Lambda$ and the parameter $a$ on the three poles discussed above. First, we discuss the $J^P = 1/2^-( B\Sigma)$ system. This pole lies on the physical $(---++)$ RS, and its position below the $ B\Sigma$ threshold together with its coupling constants to various channels are listed in Table~\ref{tab:pole_1}. The coupling constant of the $ B\Sigma(^2S_{1/2})$ channel is significantly larger than those in the other channels, indicating that this channel provides the dominant contribution and that the pole originates from a bound state in the $ B\Sigma$ single channel. 
For $a=0$, as $\Lambda$ increases, the interaction becomes stronger, and the pole moves below the threshold along the real axis on the physical sheet, reaching approximately $5$~MeV below the threshold at $\Lambda = 1.35$~GeV.
Furthermore, the pole trajectory is also affected by the parameter $a$. When $a$ increases, its real part moves along the real axis below the threshold, while its imaginary part decreases for $a<0.5$ and increases for $a>0.5$. This behavior is governed by the dynamics of the coupled-channel potential, confirming that the pole is always dominated by the $ B\Sigma$ channel.

An interesting feature of Table~\ref{tab:pole_1} is that, in some cases, the coupling to a D-wave channel exceeds that to the corresponding S-wave channel. For example, at $a=0$ and $\Lambda=1.30$ GeV, the couplings to $B^*\Lambda(^2S_{1/2})$ and $B^*\Lambda(^4D_{1/2})$ are $0.26$ and $0.63$ GeV, respectively. This behavior can be understood qualitatively from the coupled-channel dynamics and the tensor components of the OBE interaction. As shown in Eqs~\eqref{eq:poten-in-p}, the pseudoscalar- and vector-meson-exchange potentials contain tensor structures that couple partial waves with orbital angular momenta differing by two units. Consequently, the attractive S-wave channels can couple to D-wave scattering channels, generating non-negligible D-wave pole residues. The relative magnitudes of these residues are determined by the full coupled-channel interaction and the corresponding channel thresholds, and there is no general requirement that an S-wave residue must always exceed a D-wave residue. Such patterns are also observed in other coupled-channel studies of heavy-flavor molecules~\cite{Huang:2026wrg,Chen:2016heh}. Nevertheless, the comparatively large D-wave coupling in $B^*\Lambda$ should not be confused with the overall partial-wave composition of the pole. At the same parameter values, the coupling to $B\Sigma(^2S_{1/2})$ is $6.15$ GeV, considerably larger than either coupling to $B^*\Lambda$. The molecular assignment is therefore supported by the combined evidence from the single-channel attraction, the pole position, and the residue pattern, rather than by interpreting an individual coupling constant as a wave-function probability.

\begin{figure*}[ht!]
    \centering
    \begin{subfigure}[b]{0.4\textwidth}
        \includegraphics[width=\textwidth]{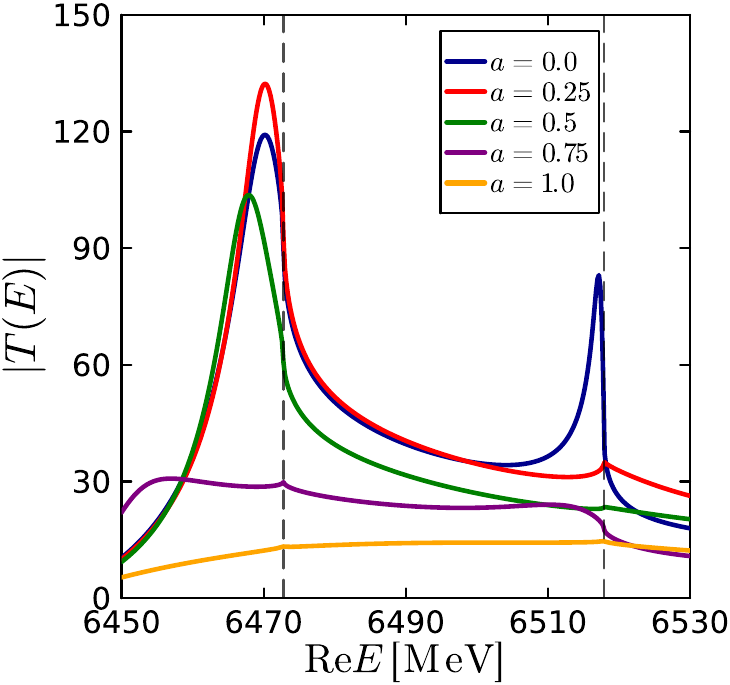}
        \caption{$ B\Sigma(^2S_{1/2}) \to B_s^0N(^2S_{1/2})$ }
        \label{fig:c3q1_a}
    \end{subfigure}
     \hfill
    \begin{subfigure}[b]{0.4\textwidth}
        \includegraphics[width=\textwidth]{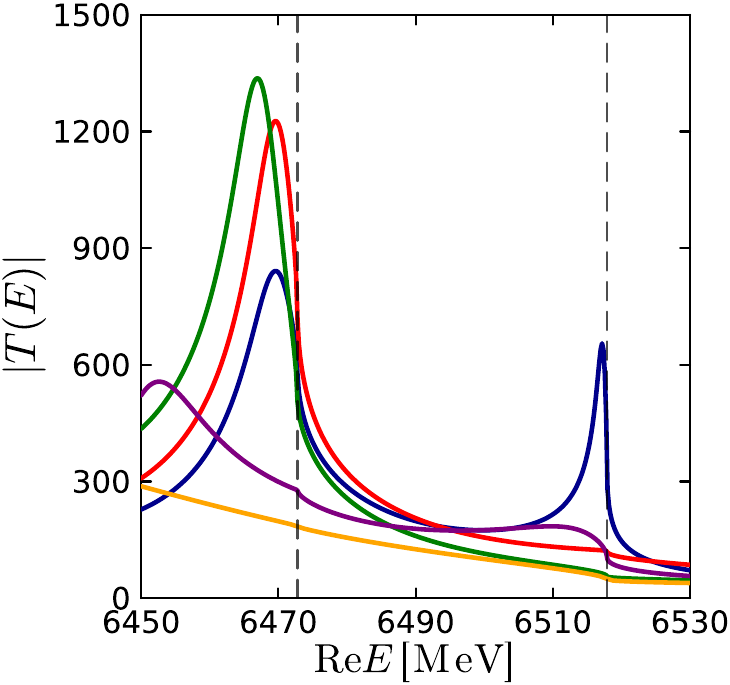}
        \caption{ $ B^*\Sigma(^2S_{1/2}) \to B_s^0N(^2S_{1/2})$ }
        \label{fig:c3q1_b}
    \end{subfigure}
    \caption{For $\Lambda = 1.3\ \text{GeV}$, scattering amplitude of the $J^P = 1/2^-$ system for $a = 0 \to 1.0$. The horizontal axis is $\operatorname{Re}(E)$, the vertical axis is $|T(E)|$. Vertical dashed lines mark the $ B\Sigma$ and $ B^*\Sigma$ thresholds at $6472.8\ \text{MeV}$ and $6517.9\ \text{MeV}$, respectively.}
    \label{fig:c3q1}
\end{figure*}

Meanwhile, the parameter $a$ influences the near-threshold behavior, as shown in Fig.~\ref{fig:c3q1}. For $a=0$, the scattering amplitude $|T(E)|$ shows a near-threshold peak on the physical real axis close to the higher threshold near $6517$ MeV, indicating the likely formation of a resonance, as illustrated in Figs.~\ref{fig:c3q1_a} and~\ref{fig:c3q1_b}. As $a$ increases, this near-threshold peak at the higher threshold disappears, demonstrating that the strength of the short-range interaction plays a decisive role in the emergence of the near-threshold virtual-state peak.

\begin{table*}[ht!]
	\centering
	\setlength{\tabcolsep}{2pt}
	\caption{Positions and absolute value of coupling constant $|g_i|$ of the pole below $ B\Sigma$ channel threshold in $J^P=1/2^-$ system as a function of $\Lambda$ and parameter $a$.}
	\label{tab:pole_1}
	\begin{ruledtabular}
	\begin{tabular}{cccccccccc}
        &  &  & \multicolumn{7}{c}{$|g_i|$ (GeV)} \\
		\cmidrule(lr){4-10}
		$a$ & $\Lambda$ (GeV) & $M-i\Gamma/2$ (MeV)& \multicolumn{1}{c}{$B_s^0 N$} & \multicolumn{1}{c}{$ B\Lambda$} & \multicolumn{2}{c}{$ B^*\Lambda$} & \multicolumn{1}{c}{$ B\Sigma$} & \multicolumn{2}{c}{$ B^*\Sigma$} \\
		\cmidrule(lr){4-4} \cmidrule(lr){5-5} \cmidrule(lr){6-7} \cmidrule(lr){8-8} \cmidrule(lr){9-10}	& & & $^2S_{1/2}$ & $^2S_{1/2}$ & $^2S_{1/2}$ & $^4D_{1/2}$ & $^2S_{1/2}$ & $^2S_{1/2}$ & $^4D_{1/2}$ \\
		\hline
	         & 1.25 & $6471.9 - i0.9$ & 1.03& 0.69&0.18&  0.73&5.56 & 4.88&3.21\\
		0.0  & 1.3  & $6471.2 - i1.6$ &1.32 & 0.71& 0.26&0.63& 6.15& 4.22& 3.09 \\
	         & 1.35 & $ 6467.9 - i5.5$ & 3.51 & 1.44&0.95&0.86&14.28  & 4.20&3.85 \\
		 \hline
	         & 1.25 &  $6471.4 - i0.7$ & 0.91 & 0.39 & 0.61&0.71&6.07&5.15&3.22\\
		0.25 & 1.3  & $6470.8 - i1.6$  &1.48 &0.57 &0.82 &0.83& 7.35&6.90& 3.57 \\
	         & 1.35 & $6469.1-i 3.6$   & 2.62 & 0.98& 1.12 & 1.08&11.15 & 10.15& 4.85\\
	     \hline
         
	         & 1.25 &  $6471.4-i0.1$ & 0.26 & 0.10&0.45&0.21&2.49&1.95 &0.68\\
	     0.5 & 1.3  & $6468.5-i0.6$    &0.47&0.19& 0.60& 0.30& 3.93&  3.20&   1.31\\
	         & 1.35 & $6466.4-i 2.2$ & 1.32&0.58 &1.40 & 0.55&8.43&  7.76&2.68\\
	     \hline
	         & 1.25 & $6463.2-i0.5$ & 0.75&0.77&1.78& 0.31&4.27&3.13&0.17\\
	    0.75 & 1.3  & $6460.3-i1.4$ &1.12&0.86&2.34&0.44&8.96 & 6.67&2.80 \\
	         & 1.35 & $6450.3-i2.3$ & 2.46 & 1.53 &2.41& 0.43&11.31& 9.16 & 4.47\\
	     \hline
	         & 1.25 & $6443.1-i0.7$ &0.06& 0.30 &0.85&0.04& 3.23& 0.75&0.62 \\
	     1.0 & 1.3  & $6442.7-i3.2$ &0.51 &3.19&7.29& 0.29& 28.84& 8.43& 4.16\\
	         & 1.35 & $6441.9-i 5.7$& 1.34 & 6.73& 8.88&0.61& 30.64& 11.05& 5.01\\
	\end{tabular}
	\end{ruledtabular}
\end{table*}

\begin{figure}[ht!]
	\centering
	\includegraphics[width=0.45\textwidth]{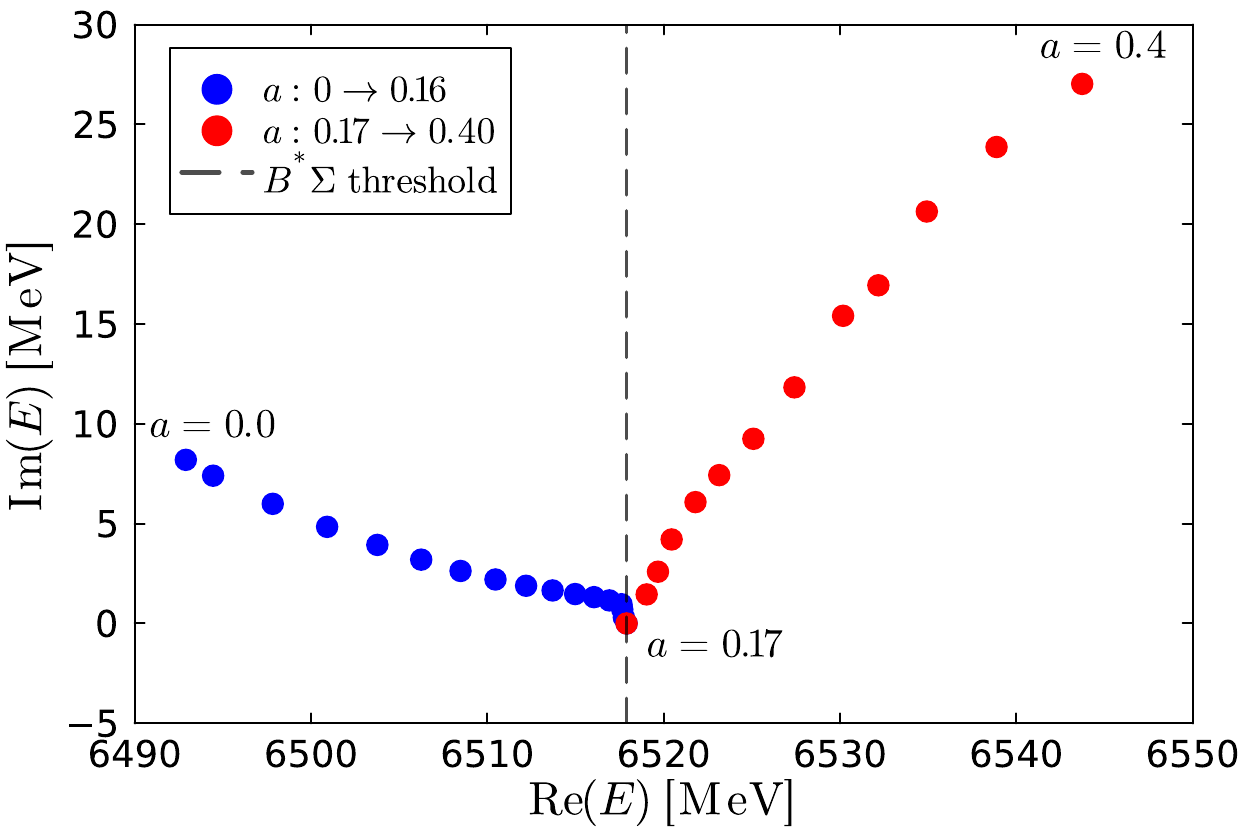}
	\caption{Trajectory of the pole below the $ B^*\Sigma$ threshold in $J^P=1/2^-$ system for $\Lambda = 1.4\ \text{GeV}$ as a function of the parameter $a$. Blue and red points correspond to the poles on $(----+)$ and $(++++-)$ RSs, respectively.}
	\label{fig:sacl.pdf}
\end{figure}

Next, we analyze the higher poles near threshold of the $ B^*\Sigma$ channel in $J^P=1/2^-$ system, selecting a cutoff $\Lambda\sim 1.3$ GeV and examining the pole behavior on the $(----+)$ RS; its position and couplings are listed in Table~\ref{tab:c5q1}.
The coupling constant of the $^2S_{1/2}$ partial wave in the $ B^*\Sigma$ channel is much larger than those in the other channels, indicating that this channel provides the dominant contribution and that the pole originates from a bound state in the $ B^*\Sigma$ single channel. For $a=0$, as $\Lambda$ increases, the interaction becomes stronger, and the pole moves below the threshold along the real axis on the physical RS, reaching about $25$~MeV below the threshold at $\Lambda = 1.4$~GeV, while its imaginary part increases from $0.2$ to $8.2$~MeV. The pole trajectory is also affected by the parameter $a$, as shown in Fig.~\ref{fig:sacl.pdf}. When $a$ increases from $0$ to $0.17$, the pole trajectory shifts rightward along the real axis toward the threshold, its imaginary part gradually decreases, and it approaches the bound-state limit, forming a near-threshold bound state, which indicates that attraction dominates. For $a > 0.17$, the pole jumps to the unphysical $(++++-)$ RS, and moves rightward, away from the threshold.

\begin{table*}[ht!]
	\centering
	\setlength{\tabcolsep}{3pt}
	\caption{ Pole positions and absolute value of coupling constant $|g_i|$ of the pole below $ B^*\Sigma$ channel threshold in $J^P=1/2^-$ system as a function of $\Lambda$ and parameter $a$.}
	\label{tab:c5q1}
    \begin{ruledtabular}
	\begin{tabular}{cccccccccc}
		& & & \multicolumn{7}{c}{$|g_i|$ (GeV)} \\
		\cmidrule(lr){4-10}
		$a$ & $\Lambda$ (GeV) & $M-i\Gamma/2$ (MeV) & \multicolumn{1}{c}{$B_s^0 N$} & \multicolumn{1}{c}{$ B\Lambda$} & \multicolumn{2}{c}{$ B^*\Lambda$} & \multicolumn{1}{c}{$ B\Sigma$} & \multicolumn{2}{c}{$ B^*\Sigma$} \\
		\cmidrule(lr){4-4} \cmidrule(lr){5-5} \cmidrule(lr){6-7} \cmidrule(lr){8-8} \cmidrule(lr){9-10}	& & & $^2S_{1/2}$ & $^2S_{1/2}$ & $^2S_{1/2}$ & $^4D_{1/2}$ & $^2S_{1/2}$ & $^2S_{1/2}$ & $^4D_{1/2}$ \\
		\hline
		& 1.3 & $ 6517.7-i0.2$ & 0.83&1.45&  0.71&  0.60&1.10& 10.32&0.04 \\
		0.0  & 1.35  & $ 6510.9-i2.8$ &0.91 &3.31& 0.42 & 1.57& 1.52&22.38&0.52 \\
		& 1.4& $ 6492.9- i 8.2$ & 0.15 & 5.53& 2.93&3.22& 1.86 & 38.75&2.88\\
	\end{tabular}
    \end{ruledtabular}
\end{table*}

\begin{figure*}[ht!]   
   \begin{subfigure}[a]{0.3\textwidth} \includegraphics[width=\textwidth]{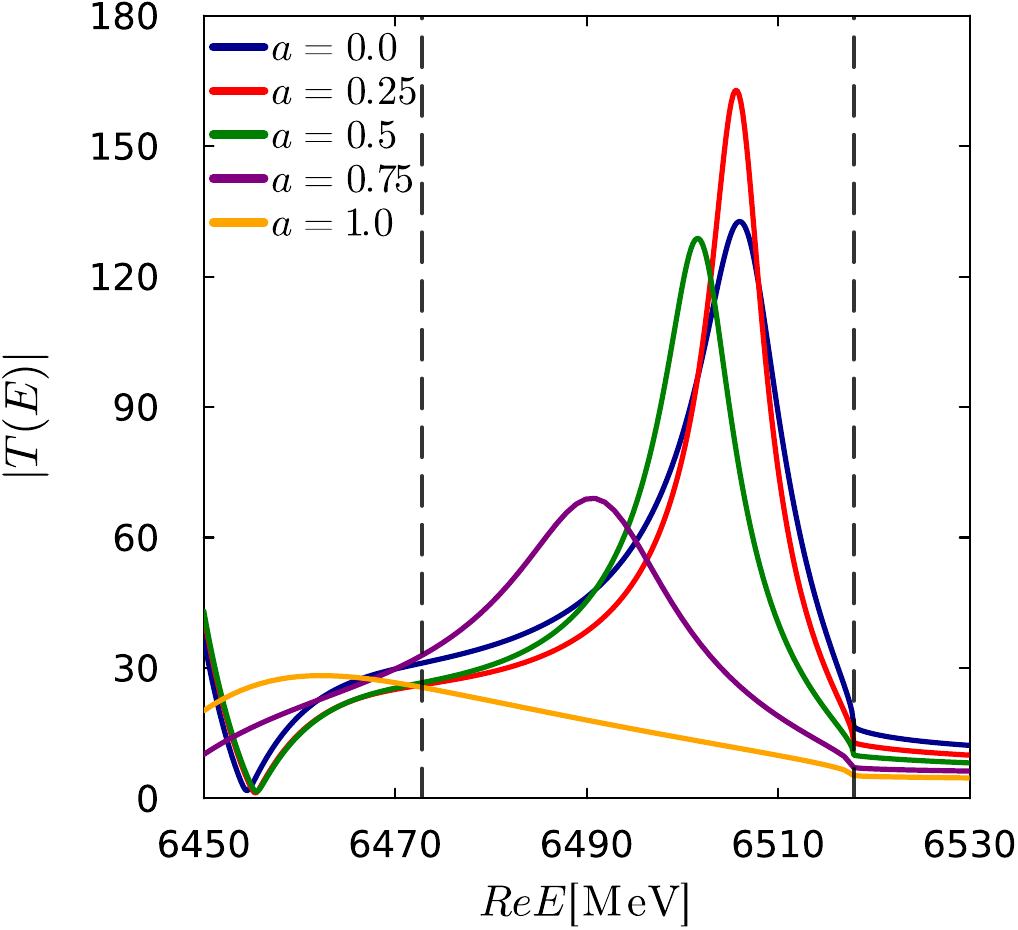}
        \caption{$ B^*\Sigma(^4S_{3/2}) \to B_s^0N(^2D_{3/2})$}
        \label{fig:plot_1}
    \end{subfigure}
    \hfill
    \begin{subfigure}[a]{0.3\textwidth}
       \includegraphics[width=\textwidth]{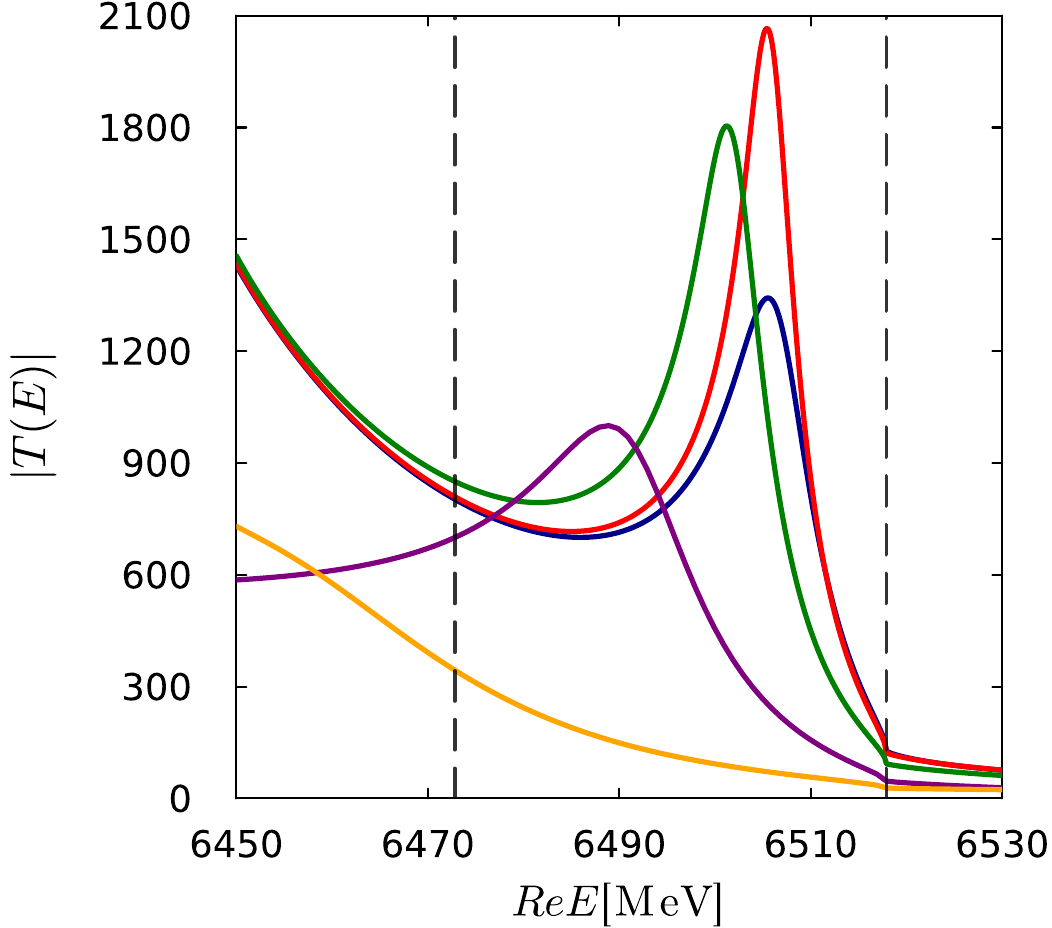}
        \caption{$ B^*\Sigma(^4S_{3/2}) \to  B\Lambda(^2D_{3/2})$}
        \label{fig:plot_2}
    \end{subfigure}
     \hfill
    \begin{subfigure}[a]{0.3\textwidth}
       \includegraphics[width=\textwidth]{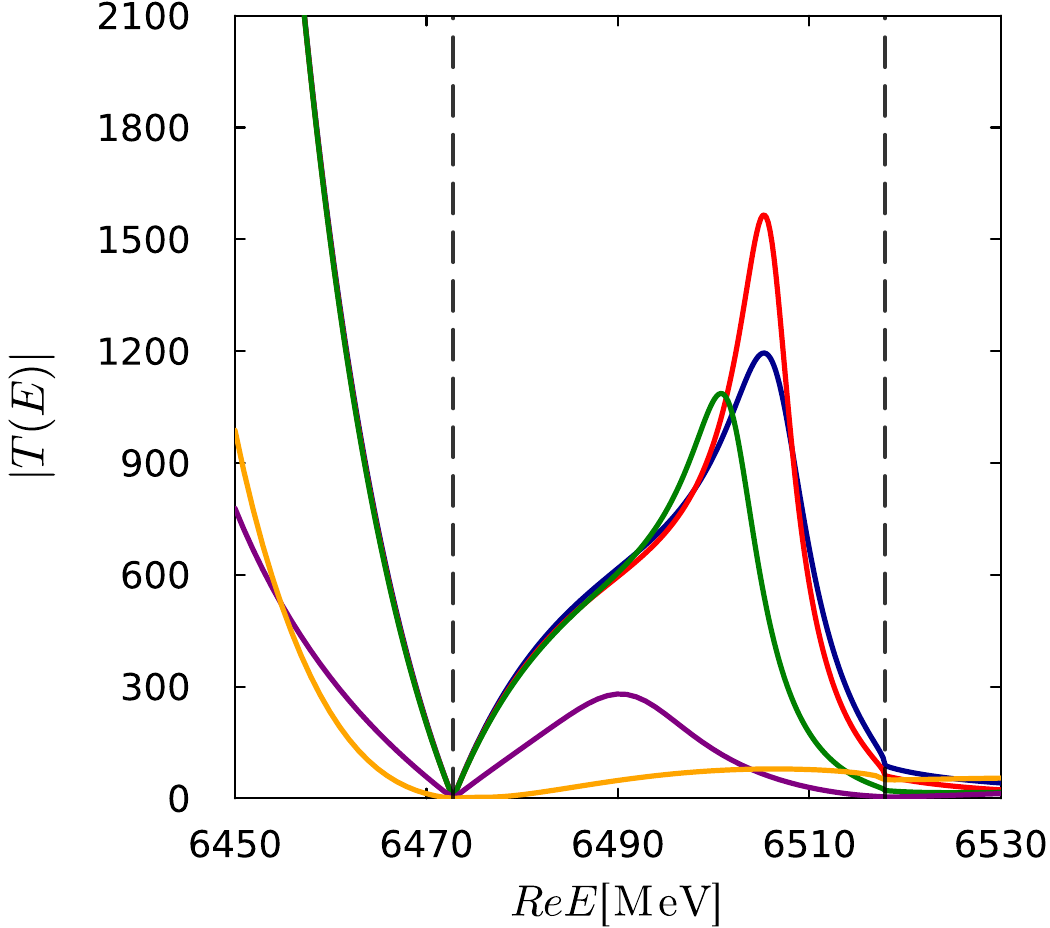}
        \caption{$ B^*\Sigma(^4S_{3/2}) \to  B^*\Lambda(^4S_{3/2})$}
        \label{fig:plot_3}
    \end{subfigure}
    \caption{For $\Lambda = 1.3\ \text{GeV}$, scattering amplitude of the $J^P = 3/2^-$ system for $a = 0 \to 1.0$. The horizontal axis is $\operatorname{Re}(E)$, the vertical axis is $|T(E)|$. The dashed vertical lines mark the $ B\Sigma$ and $ B^\ast\Sigma$  thresholds.}
    \label{c5q2_1x2.pdf}
\end{figure*}

Moreover, we analyze the $J^P = 3/2^-( B^*\Sigma)$ system. The pole lies on the $(----+)$ RS, and its properties are listed in Table~\ref{tab:c5q2}. The coupling constant of the $^4S_{3/2}$ partial wave in the $ B^*\Sigma$ channel is much larger than those in the other channels. As the interaction strengthens with increasing $\Lambda$ from $1.2$ to $1.3$~GeV, the pole moves below the threshold along the real axis on the $(----+)$ sheet, reaching approximately $8.5$~MeV below the threshold at $\Lambda = 1.3$~GeV. With a further increase of the cutoff $\Lambda$ , the pole continues to move deeper; we restrict our analysis, however, to bound states within $50$ MeV below the threshold, close to the $ B\Sigma$ threshold. Meanwhile, the dependence on the parameter $a$ is also evident in the scattering amplitudes shown in Fig.~\ref{c5q2_1x2.pdf}. As $a$ increases from $0$ to $1.0$, the pole moves below the threshold along the real axis, indicating that threshold repulsion from the $ B^*\Sigma$ channel dominates when the contact-term attraction is suppressed. At $a=0.25$, the transition $ B^*\Sigma(^4S_{3/2}) \to  B\Lambda(^2D_{3/2})$ exhibits the largest scattering amplitude as shown in Fig.~\ref{fig:plot_2}, implying that the reaction is more likely to occur and a bound state is formed.

\begin{table*}[ht!]
	\centering
	\setlength{\tabcolsep}{2pt}
	\caption{Pole positions and absolute value of coupling constant $|g_i|$ in the $J^P = 3/2^-( B^*\Sigma)$ system.}
	\label{tab:c5q2}
	\begin{ruledtabular}
	\begin{tabular}{cccccccccccc}
		&  &  & \multicolumn{9}{c}{$|g_i|$ (GeV)} \\
		\cmidrule(lr){4-12}
		$a$&$\Lambda$ (GeV) &$M-i\Gamma/2$ (MeV) & \multicolumn{1}{c}{$B_s^0 N$} & \multicolumn{1}{c}{$ B\Lambda$} & \multicolumn{3}{c}{$ B^*\Lambda$} & \multicolumn{1}{c}{$ B\Sigma$} & \multicolumn{3}{c}{$ B^*\Sigma$} \\
		\cmidrule(lr){4-4} \cmidrule(lr){5-5} \cmidrule(lr){6-8} \cmidrule(lr){9-9} \cmidrule(lr){10-12}
		& & & $^2D_{3/2}$ & $^2D_{3/2}$ & $^4S_{3/2}$ & $^2D_{3/2}$ & $^4D_{3/2}$ & $^2D_{3/2}$ & $^4S_{3/2}$ & $^2D_{3/2}$ & $^4D_{3/2}$ \\
		\hline
		& 1.2 & $6514.1-i2.4$  &0.07&2.20& 1.79&1.80 & 0.72& 2.08& 20.04 & 0.15 & 0.34 \\
		0.0   & 1.25 & $6511.0-i2.6$  &0.14 & 2.32 &  1.80 & 2.25  & 0.78& 2.15& 24.98& 0.23 &0.56\\
		& 1.3 & $ 6509.6 - i2.9$  &0.25 & 2.46 & 2.08& 2.88&0.82& 2.29 &  29.17& 0.53&1.26 \\
		\hline
		
		& 1.2 & $6513.2 -i1.9$ & 0.11 &2.07& 1.41 &0.07& 0.68&2.05 & 21.82 & 0.20 & 0.44\\
		0.25 & 1.25 & $6509.2 -i2.0$ & 0.12 & 2.32 & 1.56 & 0.08& 0.81& 2.18& 24.97&0.23& 0.56 \\
		& 1.3 &$ 6505.8 -i2.3$&0.19 & 2.38 & 1.47& 0.13 & 0.75& 2.19 &28.73 & 0.31& 0.80 \\
		\hline
		
		& 1.2 & $6509.7-i2.3$ &0.08 &2.33&1.40& 1.98 &0.87 &2.19& 24.91&0.21& 0.48\\
		0.5 & 1.25 & $6506.1 -i2.4$ & 0.12&2.37 &  1.30 & 2.23 & 0.81&2.21& 28.88 &0.27 & 0.70 \\
		& 1.3 &$ 6501.8 -i2.5$&0.18 & 2.39& 1.17 & 2.65& 0.74&2.18 &32.97& 0.35& 0.96 \\
		\hline
		
		& 1.2& $6503.8 -i3.9$ & 0.07 &2.30&0.55 &4.29 &2.55 &2.71 & 32.83 & 0.82 & 2.24\\
		0.75 & 1.25 & $ 6497.8 -i4.2$& 0.13&2.38 & 0.90 & 5.72& 0.76&2.13& 37.28 &0.37&  1.02 \\
		& 1.3 &$ 6491.4 -i 6.5$&0.17 & 2.32 &  0.74&6.57& 0.65&1.98 &  43.24& 0.52& 1.37 \\
		\hline
		
		& 1.2 & $ 6492.7-i11.9$ & 0.09&2.38&0.71 &9.90&0.75&2.03 & 46.45&0.59& 0.97\\
		1.0 & 1.25 & $ 6483.5 -i13.4 $ &0.08 &2.26& 1.17 & 12.58 &0.60&1.73& 60.71 &2.01& 3.20\\
		& 1.3 &$ 6473.1-i14.2$&0.12& 1.60& 3.33 &13.08& 0.58&1.89&  69.50 & 3.73& 3.43\\ 
		
	\end{tabular}
	\end{ruledtabular}
\end{table*}

As a benchmark, we calculate the the partial decay widths of these three poles when $a=0$, and results are collected in Table~\ref{tab:res-a-0.58}. For the $J^P=1/2^-( B\Sigma)$ pole, the dominant decay is found in the lowest open channel, indicating that this state is most likely to be observed through the $B_s^0N$ spectrum, with the $ B\Lambda$ and $ B^*\Lambda$ channels giving smaller but still relevant contributions. For the $J^P=1/2^-( B^*\Sigma)$ pole, the state is generated mainly by the near-threshold $ B^*\Sigma$ channel as shown in Table~\ref{tab:c5q1} , its decay width is concentrated in the $ B\Lambda$ channel. For the $J^P=3/2^-( B^*\Sigma)$ pole, the $B_s^0N$ decay is strongly suppressed, consistent with its  $D$-wave partial decay channel, while the dominant decay channels are distributed among the $ B\Lambda$, and $ B^*\Lambda$ channels. These results show that the predicted states are relatively narrow and may be experimentally accessible as near-threshold structures rather than broad resonances. Therefore, future searches should focus not only on a single invariant mass distribution, but also on correlated coupled-channel line-shape analyses involving $B_s^0N$, $ B\Lambda$, and $ B^*\Lambda$ final states.

\begin{table*}[ht!]
\centering
\caption{Partial decay widths $\Gamma_i$ (in MeV) for the $1/2^-( B\Sigma)$,  $1/2^-( B^*\Sigma)$ ,and  $3/2^-( B^*\Sigma)$ channels when $a=0$. The pole positions are given as $M - i\Gamma/2$ (in MeV). For the $1/2^-( B\Sigma)$ pole, the partial decay widths are listed for $B_s^0N(^2S_{1/2})$, $ B \Lambda(^2S_{1/2})$, $ B^*\Lambda(^2S_{1/2})$, and $ B^*\Lambda(^4D_{1/2})$. For the $1/2^-( B^*\Sigma)$ pole, the additional  $ B\Sigma(^2S_{1/2})$ channel is included. For the $3/2^-( B^*\Sigma)$ pole, the listed channels are $B_s^0N(^2D_{3/2})$, $ B\Lambda(^2D_{3/2})$, $ B^*\Lambda(^4S_{3/2}), B^*\Lambda(^2D_{3/2}), B^*\Lambda(^4D_{3/2})$, and $ B\Sigma(^2D_{3/2})$.}\label{tab:res-a-0.58}\setlength{\tabcolsep}{1pt}
\begin{ruledtabular}
\begin{tabular}{c|c|cccc|c|ccccc|c|ccccccc}
\multirow{2}{*}{$\Lambda$} & \multicolumn{5}{c|}{$J^P=1/2^-( B\Sigma)$} & \multicolumn{6}{c|}{$J^P=1/2^-( B^*\Sigma)$} & \multicolumn{7}{c}{$J^P=3/2^-( B^*\Sigma)$} \\
& $M-i\Gamma/2$ & \multicolumn{4}{c|}{$\Gamma_i$} & $M-i\Gamma/2$ & \multicolumn{5}{c|}{$\Gamma_i$} & $M-i\Gamma/2$ & \multicolumn{6}{c}{$\Gamma_i$} \\
\hline
$ 1.3  $ & $6471.2 -i1.6$ & $2.0$& $1.2$ & $0.0$& $0.2$& $ 6517.7-i0.2$  & $  0.1 $ & $  0.2 $ & $  0.0 $ & $  0.0 $ & $  0.1 $ & $ 6509.6 -i2.9$ & $  0.0$ & $  2.0  $ & $ 1.1$ & $ 2.8  $ & $ 0.2 $ & $0.8$\\
$ 1.35 $ & $ 6467.9 - i5.5$ &$8.5$ & $1.1$ &$0.3$ &$0.2$ & $  6510.9-i2.8  $ & $  0.1 $ & $  2.3 $ & $ 0.3$ & $  1.9 $ & $  2.1$ & $  6506.5-i3.4 $ & $  0.1$ & $  2.2$ & $ 0.1$ & $   3.4  $ & $  0.4$ & $ 1.1 $\\
$ 1.4  $ & $ 6462.5-i7.5  $ & $ 12.4 $ & $  1.1 $ & $  1.0 $&$ 0.3$ & $ 6492.9- i 8.2$& $ 0.1 $ & $  12.2$ & $  1.6$ & $  3.1 $ & $  1.1 $ & $6495.0-i5.1 $ & $  0.6$ & $ 3.0  $ & $ 0.4 $ & $  4.8  $ & $ 0.5  $ & $ 1.2 $	
\end{tabular}
\end{ruledtabular}
\end{table*}

\subsection{Extended analysis involving the $B\Sigma^*$ and $B^*\Sigma^*$ channels}
In the previous analysis, the coupled-channel dynamics of five channels consisted  of the combinations of $B^{(*)}$ meson and SU(3) octet baryons($N,\Sigma,\Lambda$) are considered. For a more complete description of the coupled-channel dynamics, it is also necessary to examine the channels involving the spin-$3/2$ decuplet baryon $\Sigma^*$, namely $B\Sigma^*$ and $B^*\Sigma^*$.
These two channels carry the same quark configuration as the five channels considered above and can therefore couple to them through light-meson exchange. Their thresholds are $W_{B\Sigma^*}=6662.5$ MeV and $W_{B^*\Sigma^*}=6707.6$ MeV, which are about $140$--$190$ MeV higher than the $B^*\Sigma$ threshold. We thus extend the original five-channel basis to the seven-channel basis $B_s^0N, B\Lambda, B^*\Lambda, B\Sigma, B^*\Sigma, B\Sigma^*$ and $B^*\Sigma^*$. 
The Riemann sheets quoted below are specified according to this channel ordering. The interaction potentials involving the newly included $B\Sigma^*$ and $B^*\Sigma^*$ channels are derived in Appendix~\ref{app:decuplet-pot}. In constructing these potentials, the heavy-meson vertices are the same as those used in the previous five-channel calculation, while the light-baryon vertices are extended to include the diagonal decuplet interactions and the octet-decuplet transition interactions. The same monopole form factor and the same prescription for the short-range $\delta(r)$ term are used as in the original calculation. For the newly added channels, we consider the negative-parity states with $J^P=1/2^-$ and $3/2^-$. The corresponding partial-wave components are

\begin{itemize}
	\item $J^P=1/2^-$:\\
    $^4D_{1/2}(B\Sigma^*)$,
    $[^2S_{1/2}, ^4D_{1/2}, ^6D_{1/2}](B^*\Sigma^*)$,
	\item $J^P=3/2^-$: \\
	$[^4S_{3/2}, ^4D_{3/2}](B\Sigma^*)$, 
    $[^4S_{3/2}, ^2D_{3/2}, ^4D_{3/2}, ^6D_{3/2}](B^*\Sigma^*)$.
\end{itemize}

We first examine how the inclusion of the $B\Sigma^*$ and $B^*\Sigma^*$ channels affects the three poles associated with the $B\Sigma$ and $B^*\Sigma$ thresholds. The results for $a=0$ are summarized in Table~\ref{tab:pole-mass}. Compared with the five-channel calculation in the Tables~\ref{tab:pole_1}, \ref{tab:c5q1} and \ref{tab:c5q2}, the pole positions are shifted only moderately. The lower $J^P=1/2^-$ pole remains located slightly below the $B\Sigma$ threshold and is still dominated by the $B\Sigma$ component. The two poles associated with the $B^*\Sigma$ threshold also remain close to their original positions. This indicates that the higher $B\Sigma^*$ and $B^*\Sigma^*$ channels do not qualitatively change the molecular interpretation of the near-threshold $B\Sigma$ and $B^*\Sigma$ poles, although they provide additional coupled-channel corrections. 

\begin{table*}[ht!]
\caption{Pole positions $M-i\Gamma/2$ for the three original $B\Sigma$- and $B^*\Sigma$-dominated poles after including the $B\Sigma^*$ and $B^*\Sigma^*$ channels as the cutoff varies. The short-range parameter is fixed at $a=0$.}
\label{tab:pole-mass}
\begin{ruledtabular}
\begin{tabular}{cc|cc|cc}
\multicolumn{2}{c|}{$J^P=1/2^-( B\Sigma)$}&\multicolumn{2}{c|}{$J^P=1/2^-( B^*\Sigma)$}&\multicolumn{2}{c}{$J^P=3/2^-( B^*\Sigma)$}\\
$\Lambda$(GeV) & $M-i\Gamma/2$(MeV)& $\Lambda$ & $M-i\Gamma/2$ & $\Lambda$ & $M-i\Gamma/2$ \\
\hline
$1.25$  & $6471.2-i0.6$  & $1.3$ & $6517.1-i0.3$  & $1.2$ & $6513.8-i2.2$ \\
$1.3$   & $6470.9-i2.1$  & $1.35$ & $6511.3-i2.4$ & $1.25$ & $6509.4-i2.5$ \\
$1.35$  & $6467.1-i6.5$  & $1.4$ & $6493.5-i8.4$  & $1.3$ & $6507.3-i3.7$ 
\end{tabular}
\end{ruledtabular}%
\end{table*}

Next, we discuss the pole structures associated with the newly included $B\Sigma^*$ and $B^*\Sigma^*$ channels. Three near-threshold poles are found in the extended coupled-channel calculation, one is located near the $B\Sigma^*$ threshold in the $J^P=3/2^-$ system, and denoted with $J^P=3/2^-(B\Sigma^*)$ pole. The other two are found  near the $B^*\Sigma^*$ threshold in the $J^P=1/2^-$ and $3/2^-$ systems, which are denoted as $J^P=1/2^-(B^*\Sigma^*)$ and $J^P=3/2^-(B^*\Sigma^*)$ poles, respectively. The corresponding pole positions are listed in Table~\ref{tab:pole-mass_1}.

For the $J^P=3/2^-(B\Sigma^*)$ pole, in the considered cutoff region, its mass lies in the range $6.65$--$6.66$ GeV for $a=0$, with $\Gamma/2=1.97$--$4.40$ MeV. When the parameter $a$ is increased, the pole moves closer to the threshold and its width decreases. This behavior indicates that the short-range $\delta(r)$ term enhances the binding of this state. The couplings listed in Table~\ref{tab:combined} show that the $B\Sigma^*({}^4S_{3/2})$ component gives the dominant contribution, with $|g_i|$ much larger than those of most other channels. This supports the interpretation of this pole as a predominantly $B\Sigma^*$ molecular state, their coupling to other channels($B_s^0N, B^{(*)}\Lambda$) are relatively small and unlisted here.

Another pole in the $J^P=3/2^-$ system is appeared near $B^*\Sigma^*$ threshold system. The dominant coupling is to the $B^*\Sigma^*({}^4S_{3/2})$ component, as shown in Table~\ref{tab:combined}. For $a=0$, the pole moves farther below threshold and its width increases as $\Lambda$ increases, which is typical for a near-threshold molecular pole generated by an attractive interaction. When $a$ is increased, the pole moves closer to threshold and becomes narrower. The pole survives for moderate values of $a$, but it is not found at the RS ($------+$) connected to the physical real energy axis when the contact term is fully removed. This indicates that both the long-range meson-exchange interaction and the short-range contribution play relevant roles in this channel.

In the $J^P=1/2^-$ system, a pole is found near the $B^*\Sigma^*$ threshold. For $a=0$, the pole moves from $6703.0-i0.7$ MeV to $6675.6-i2.8$ MeV as the cutoff increases from $1.05$ to $1.15$ GeV, corresponding to an increasing binding energy and width. The residue analysis in Table~\ref{tab:combined} shows that this pole couples dominantly to the $B^*\Sigma^*({}^2S_{1/2})$ channel. However, its existence is sensitive to the short-range interaction. The pole survives at the RS($------+$) connected to physical real energy axis for $a=0$ and $a=0.25$, and it moves to other RSs for $a\geq0.5$ in the considered cutoff region. Therefore, this $J^P=1/2^-(B^*\Sigma^*)$ pole should be regarded as more model dependent.

\begin{table*}[htbp]
\centering
\caption{Pole positions $M-i\Gamma/2$ for the $\Sigma^*$-related molecular candidates in the extended seven-channel calculation. The $J^P=3/2^-(B\Sigma^*)$ pole lies on the $(-----++)$ sheet, while the $J^P=1/2^-(B^*\Sigma^*)$ and $J^P=3/2^-(B^*\Sigma^*)$ poles lie on the $(------+)$ sheet. The symbol ``$\cdots$'' means that the poles move to other RSs.}
\label{tab:pole-mass_1}
\begin{ruledtabular}
\begin{tabular}{c|cc|cc|cc}
	& \multicolumn{2}{c|}{$J^P=1/2^-( B\Sigma^*)$}&\multicolumn{2}{c|}{$J^P=1/2^-( B^*\Sigma^*)$}&\multicolumn{2}{c}{$J^P=3/2^-( B^*\Sigma^*)$}\\
$a$ & $\Lambda$(GeV) & $M-i\Gamma/2$(MeV) & $\Lambda$ & $M-i\Gamma/2$ & $\Lambda$ & $M-i\Gamma/2$ \\
\hline
 & $1.22$  & $6660.1-i1.9$  &  $1.05$ & $6703.0-i0.7$   & $1.15$ & $6705.3-i5.9$ \\
0.0 & $1.28$  & $6655.6-i3.5$  &  $1.10$ & $6692.9-i1.6$ & $1.20$ & $6699.3-i10.2$ \\
& $1.32$  & $6649.2-i4.4$ & $1.15$ & $6675.6-i2.8$  & $1.25$ & $6689.5-i15.1$ \\
\hline
 & $1.22$  & $6660.9-i0.7$  & $1.05$ & $6707.6-i0.0$  & $1.15$ & $6707.5-i1.3$ \\
0.25 & $1.28$  & $6657.7-i1.7$  & $1.10$ & $6706.1-i0.2$ & $1.20$ &$6705.5-i3.2$  \\
& $1.32$  &$6654.5-i2.4$   & $1.15$ &$ 6700.8-i0.7$  & $1.25$ &$6701.4-i5.3$ \\
\hline
 & $1.22$  & $6661.3-i0.4$  & $1.05$ & $\cdots$  & $1.15$ & $\cdots$  \\
0.5 & $1.28$  &$6658.6-i0.8$  & $1.10$ &$\cdots$& $1.20$ & $6707.2-i0.4$ \\
& $1.32$  & $6655.7-i1.2$  & $1.15$ &  $\cdots$ & $1.25$ & $6706.7-i1.2$ \\
\hline
 & $1.22$  & $6661.5-i0.2$  & $1.05$ &  $\cdots $  & $1.15$ & $\cdots$ \\
1.0 & $1.28$  & $6657.3-i0.5$  & $1.10$ & $\cdots$   & $1.20$ & $\cdots$\\
& $1.32$  & $6656.2-i0.6$  & $1.15$ &  $\cdots$    & $1.25$ &$\cdots$\\
\end{tabular}%
\end{ruledtabular}
\end{table*}

\begin{table*}[ht!]
\centering
\caption{Pole positions $M-i\Gamma/2$ and absolute values of pole couplings $|g_i|$ for the $\Sigma^*$-related poles in the extended seven-channel calculation when $a=0$. The upper block corresponds to the $J^P=1/2^-(B^*\Sigma^*)$ pole, while the lower block corresponds to the $J^P=3/2^-(B\Sigma^*)$ and $J^P=3/2^-(B^*\Sigma^*)$ poles.}
\label{tab:combined}
\begin{ruledtabular}
\begin{tabular}{ccc*{10}{c}}
&&&\multicolumn{7}{c}{$|g_i|$ (GeV)}&&& \\ 
\cmidrule(lr){1-13}
 & $\Lambda$ (GeV) & $M-i\Gamma/2$ (MeV)
& \multicolumn{1}{c}{$B\Sigma$} 
& \multicolumn{2}{c}{$B^*\Sigma$} & \multicolumn{1}{c}{$B\Sigma^*$} 
& \multicolumn{3}{c}{$B^*\Sigma^*$} & \multicolumn{3}{c}{} \\
\cmidrule(lr){4-4} \cmidrule(lr){5-6} \cmidrule(lr){7-7} \cmidrule(lr){8-10}
$J^P=1/2^-$& & & $^2S_{1/2}$ & $^2S_{1/2}$ & $^4D_{1/2}$ & $^4D_{1/2}$ 
& $^2S_{1/2}$ & $^4D_{1/2}$ & $^6D_{1/2}$ & & & \\
\hline
                         & 1.05 & $6703.1-i0.7$  & 0.9 & 0.5 & 0.2 & 0.4 & 16.2 & 0.0  & 0.0 & & & \\
$B^*\Sigma^*$ & 1.10 & $6692.9-i1.6$  & 1.4 & 0.8 & 0.3 & 0.3 & 25.2 & 0.0 & 0.0 & & & \\
                         & 1.15 & $6675.6-i2.8$  & 2.0 & 1.1 & 0.5 & 0.1 & 33.7 & 0.1 & 0.1 & & & \\
\hline
&&&\multicolumn{10}{c}{$|g_i|$ } \\
\cmidrule(lr){1-13}
& $\Lambda$  & $M-i\Gamma/2$  
& \multicolumn{1}{c}{$B\Sigma$} 
& \multicolumn{3}{c}{$B^*\Sigma$} & \multicolumn{2}{c}{$B\Sigma^*$} 
& \multicolumn{4}{c}{$B^*\Sigma^*$} \\
\cmidrule(lr){4-4} \cmidrule(lr){5-7} \cmidrule(lr){8-9} \cmidrule(lr){10-13}
$J^P=3/2^-$& & & $^2D_{3/2}$ 
& $^4S_{3/2}$ & $^2D_{3/2}$ & $^4D_{3/2}$ 
& $^4S_{3/2}$ & $^4D_{3/2}$ 
& $^4S_{3/2}$ & $^2D_{3/2}$ & $^4D_{3/2}$ & $^6D_{3/2}$ \\
\hline
                       & 1.22 & $6660.1-i1.9$  & 0.4 & 2.6 & 0.9 & 0.2 & 15.0 & 0.0 & 11.1 & 0.9 & 2.4 & 3.2 \\
$B\Sigma^*$ & 1.28 & $6655.6-i3.5$  & 0.6 & 3.5 & 1.1 & 0.3 & 18.1 & 0.0 & 16.2 & 1.0 & 2.5 & 3.3 \\
                       & 1.32 & $6649.2-i4.4$  & 0.8 & 4.2 & 1.3 & 0.4 & 23.5 & 0.1 & 18.9 & 1.1 & 2.7 & 3.6 \\
\hline
                         & 1.15 & $6705.3-i5.9$   & 0.5 & 3.4 & 0.8 & 1.5 & 4.2 & 0.9 & 18.7 & 0.0 & 0.1 & 0.1 \\
$B^*\Sigma^*$ & 1.20 & $6699.3-i10.2$  & 0.7 & 4.4 & 1.0 & 1.9 & 4.8 & 0.9 & 24.5 & 0.1 & 0.2 & 0.3 \\
                         & 1.25 & $6689.5-i15.1$  & 1.0 & 5.4 & 1.2 & 2.5 & 5.5 & 0.8 & 30.4 & 0.2 & 0.6 & 0.9 \\
\end{tabular}
\end{ruledtabular}
\end{table*}

\begin{figure*}[ht!]
	\centering
	\begin{subfigure}[a]{0.48\textwidth}
\includegraphics[width=\textwidth]{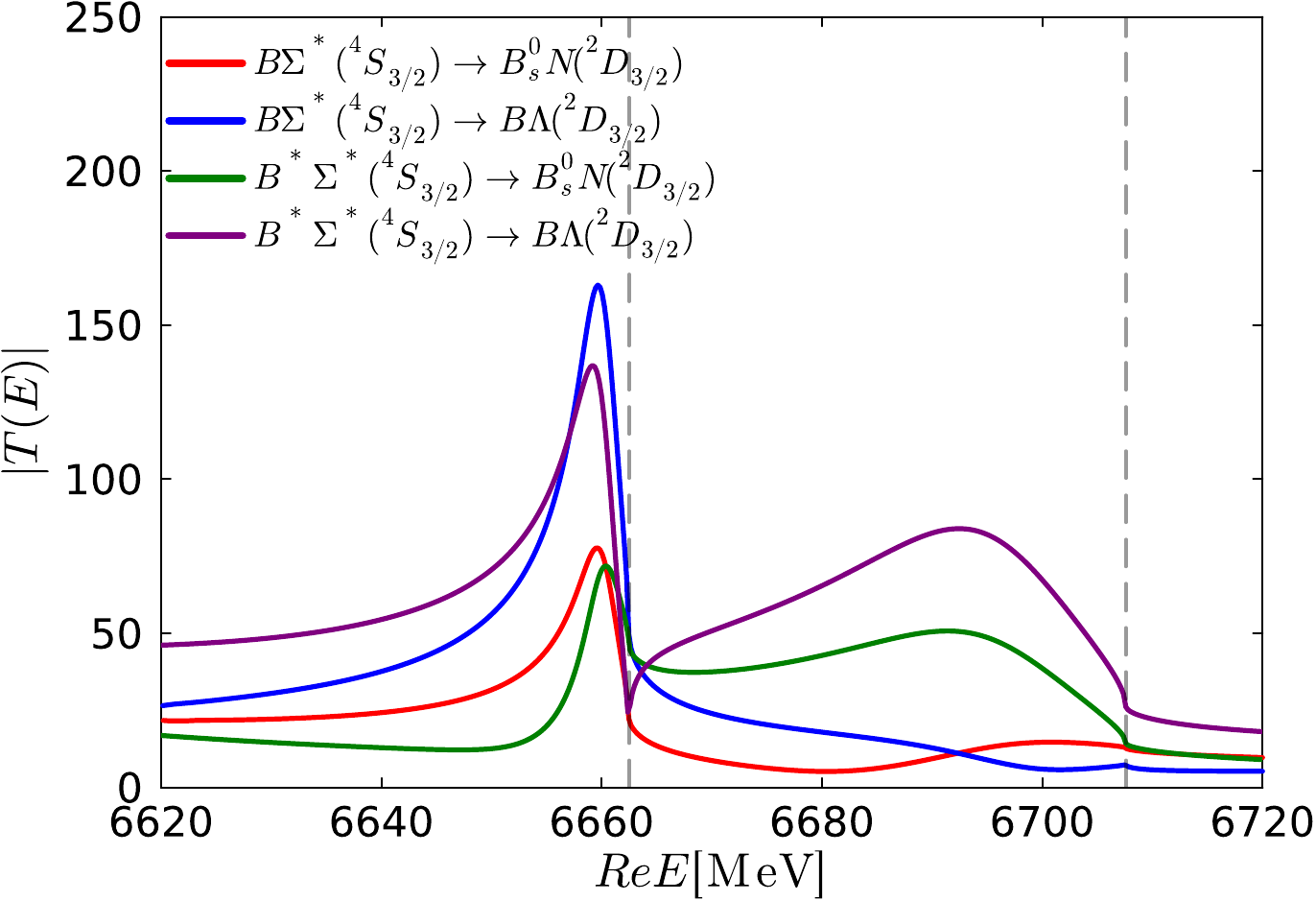}
		\caption{}
		\label{fig:3/2_c6q2}
	\end{subfigure}
	\hfill
\begin{subfigure}[a]{0.48\textwidth}
\includegraphics[width=\textwidth]{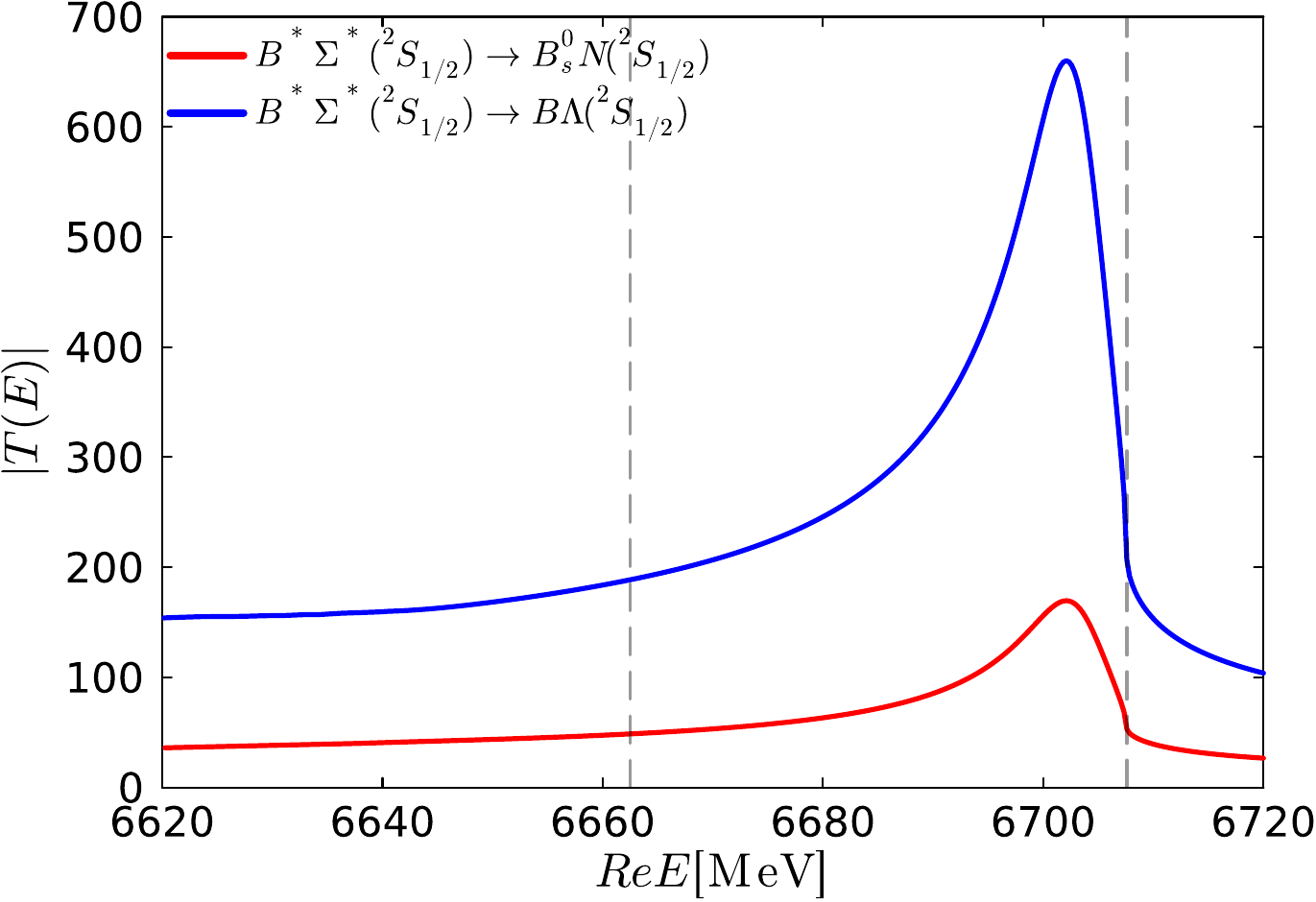}
		\caption{}
		\label{fig:c7q1}
	\end{subfigure}
    
	\caption{Moduli of the scattering amplitudes for several transition processes in the extended seven-channel calculation with $a=0$. Panel (a) shows the $J^P=3/2^-$ amplitudes for $\Lambda=1.22$ GeV, while panel (b) shows the $J^P=1/2^-$ amplitudes for $\Lambda=1.05$ GeV. The vertical dashed lines denote the $B\Sigma^*$ and $B^*\Sigma^*$ thresholds.}
	\label{fig:three poles}
\end{figure*}

The scattering amplitudes for transitions from the dominant components of these poles to the lower $B_s^0N$ and $B\Lambda$ channels are shown in Fig.~\ref{fig:three poles}, where the short-range parameter is fixed at $a=0$. For the $J^P=3/2^-$ sector with $\Lambda=1.22$ GeV, the pole near the $B\Sigma^*$ threshold produces clear peak structures in the transition amplitudes of $B^{(*)}\Sigma^*({}^4S_{3/2})\to B_s^0N({}^2D_{3/2})$ and $B^{(*)}\Sigma^*({}^4S_{3/2})\to B\Lambda({}^2D_{3/2})$ transitions.
The higher pole near the $B^*\Sigma^*$ threshold gives a broader enhancement in the corresponding transitions from the dominant $B^*\Sigma^*({}^4S_{3/2})$ component to the lower $B_s^0N$ and $B\Lambda$ channels. In contrast, some other transition amplitudes remain rather flat around the $B^*\Sigma^*$ threshold, reflecting their weaker couplings to this pole. For the $J^P=1/2^-$ sector with $\Lambda=1.05$ GeV, the pole near the $B^*\Sigma^*$ threshold produces visible near-threshold enhancements in the $B^{(*)}\Sigma^*({}^2S_{1/2})\to B_s^0N ({}^2S_{1/2})$ and $B^{(*)}\Sigma^*({}^2S_{1/2})\to B\Lambda ({}^2S_{1/2})$ transition amplitudes. These line-shape structures provide possible experimental signatures of the $\Sigma^*$-related molecular candidates.

The partial decay widths of the three $\Sigma^*$-related poles are calculated for several cutoff values with the short-range parameter fixed at $a=0$, and the results are listed in Table~\ref{tab:partial-decay-10}. For the $J^P=1/2^-(B^*\Sigma^*)$ pole, the total width is about $1.4$--$5.6$ MeV in the cutoff region $\Lambda=1.05$--$1.15$ GeV. The dominant partial decay widths are found in the $B\Lambda$ and $B\Sigma$ channels, while the contributions from $B_s^0N$ and $B^*\Lambda$ are smaller. This suggests that the $B\Lambda$ and $B\Sigma$ invariant-mass distributions are main channels for searching for this $J^P=1/2^-$ state.
For the $J^P=3/2^-(B\Sigma^*)$ pole, the total width is about $3.8$--$8.8$ MeV for $\Lambda=1.22$--$1.32$ GeV. Its decay is dominated by the $B^*\Sigma({}^4S_{3/2})$ channel, with a sizable contribution from the $B^*\Lambda({}^4S_{3/2})$ channel. 
For the $J^P=3/2^-(B^*\Sigma^*)$ pole, the total width increases from about $11.8$ MeV to $30.2$ MeV as the cutoff increases from $1.15$ to $1.25$ GeV. This pole is therefore broader than the other two $\Sigma^*$-related states. Its partial decay widths are mainly concentrated in the $B^*\Sigma({}^4S_{3/2})$ and $B\Sigma^*({}^4S_{3/2})$ channels, while the lower $B_s^0N$ and $B\Lambda$ channels provide only small contributions.

\begin{table*}[htbp!]
\centering
\caption{Partial decay widths $\Gamma_i$ in units of MeV for the $J^P=1/2^-(B^*\Sigma^*)$, $J^P=3/2^-(B\Sigma^*)$, and $J^P=3/2^-(B^*\Sigma^*)$ poles with $a=0$. The pole positions are given as $M-i\Gamma/2$ in units of MeV.}
\label{tab:partial-decay-10}
\begin{ruledtabular}
\begin{tabular}{ccc*{11}{c}}
&&&\multicolumn{7}{c}{$\Gamma_i$(GeV)} &&&\\
\cmidrule(lr){1-11}
& $\Lambda$(GeV) & $M-i\Gamma/2$(MeV)
& \multicolumn{1}{c}{$B_s^0N$} & \multicolumn{1}{c}{$B\Lambda$} 
& \multicolumn{2}{c}{$B^*\Lambda$} & \multicolumn{1}{c}{$B\Sigma$} 
& \multicolumn{2}{c}{$B^*\Sigma$} &  \multicolumn{1}{c}{$B\Sigma^*$}  \\
\cmidrule(lr){4-4} \cmidrule(lr){5-5} \cmidrule(lr){6-7} \cmidrule(lr){8-8} \cmidrule(lr){9-10} \cmidrule(lr){11-11} 
$J^P=1/2^-$& & & $^2S_{1/2}$ & $^2S_{1/2}$ & $^2S_{1/2}$ & $^4D_{1/2}$ & $^2S_{1/2}$ & $^2S_{1/2}$ & $^4D_{1/2}$ & $^4D_{1/2}$  \\
\hline
              & 1.05 & $6703.1-i0.7$  & 0.1 & 0.5 & 0.2 & 0.0 &0.5 & 0.1 & 0.0 & 0.0 \\
$B^*\Sigma^*$ & 1.10 & $6692.9-i1.6$  & 0.2 & 1.0 & 0.4 & 0.1 &1.2 & 0.3 & 0.1 & 0.0  \\
              & 1.15 & $6675.9-i2.8$  & 0.5 & 1.5 & 0.5 & 0.1 & 2.3& 0.6 & 0.1 & 0.0 \\
 \hline
 &&&\multicolumn{8}{c}{$\Gamma_i$}&& \\
\cmidrule(lr){1-12}
 & $\Lambda$ & $M-i\Gamma/2$ 
& \multicolumn{1}{c}{$B_s^0N$} & \multicolumn{1}{c}{$B\Lambda$} 
& \multicolumn{3}{c}{$B^*\Lambda$} & \multicolumn{1}{c}{$B\Sigma$} 
& \multicolumn{3}{c}{$B^*\Sigma$}  \\
\cmidrule(lr){4-4} \cmidrule(lr){5-5} \cmidrule(lr){6-8} \cmidrule(lr){9-9} \cmidrule(lr){10-12} 
$J^P=3/2^-$& & & $^2D_{3/2}$ & $^2D_{3/2}$ & $^4S_{3/2}$ & $^2D_{3/2}$ & $^4D_{3/2}$ & $^2D_{3/2}$ & $^4S_{3/2}$ & $^2D_{3/2}$ & $^4D_{3/2}$ \\
\hline
            & 1.22 & $6660.1-i1.9$  & 0.0 & 0.1 & 0.5 & 0.0 & 0.2 & 0.1 & 3.1 & 0.4 & 0.0 \\
$B\Sigma^*$ & 1.28 & $6655.6-i3.5$  & 0.0 & 0.1 & 0.9 & 0.1 & 0.3 & 0.2 & 5.8 & 0.5 & 0.0   \\
            & 1.32 & $6649.2-i4.4$  & 0.1 & 0.2 & 1.1 & 0.1 & 0.3 & 0.3 & 6.2 & 0.7 & 0.0  \\
 \hline
 &&&\multicolumn{11}{c}{$\Gamma_i$} \\
\cmidrule(lr){1-14}
& $\Lambda$ & $M-i\Gamma/2$ 
& \multicolumn{1}{c}{$B_s^0N$} & \multicolumn{1}{c}{$B\Lambda$} 
& \multicolumn{3}{c}{$B^*\Lambda$} & \multicolumn{1}{c}{$B\Sigma$} 
& \multicolumn{3}{c}{$B^*\Sigma$} & \multicolumn{2}{c}{$B\Sigma^*$} \\
\cmidrule(lr){4-4} \cmidrule(lr){5-5} \cmidrule(lr){6-8} \cmidrule(lr){9-9} \cmidrule(lr){10-12} \cmidrule(lr){13-14}
$J^P=3/2^-$& & & $^2D_{3/2}$ & $^2D_{3/2}$ & $^4S_{3/2}$ & $^2D_{3/2}$ & $^4D_{3/2}$ & $^2D_{3/2}$ & $^4S_{3/2}$ & $^2D_{3/2}$ & $^4D_{3/2}$ & $^4S_{3/2}$ & $^4D_{3/2}$ \\
\hline
               & 1.15 & $6705.3-i5.9$   & 0.0 & 0.1 & 0.1 & 0.2 & 0.5 & 0.1 & 6.1 & 0.4 & 1.3 & 5.0 & 0.2 \\
$B^*\Sigma^*$  & 1.20 & $6699.3-i10.2$  & 0.1 & 0.2 & 0.2 & 0.4 & 0.7 & 0.3 & 10.2 & 0.5 & 2.1 & 6.5 & 0.2 \\
               & 1.25 & $6689.5-i15.1$  & 0.1 & 0.4 & 0.4 & 0.6 & 0.9 & 0.6 & 14.8& 0.8& 3.1 & 6.8 &0.1\\
\end{tabular}
\end{ruledtabular}
\end{table*}

Before closing this Section, we briefly comment on possible experimental signatures of the three pole structures obtained above. Unlike the hidden-charm pentaquarks $P_c$ and $P_{cs}$, which are readily observed in the weak decays of the $\Lambda_b^0$ and $\Xi_b^-$ baryons, the hidden-bottom $P_{ \bar b s}$ pentaquarks predicted in this work possess masses higher than $6.44$ GeV. Because these masses exceed those of all established ground-state bottom hadrons (e.g., $m_{\Lambda_b} \approx 5620$ MeV, $m_{\Xi_b} \approx 5797$ MeV, $m_{B_c} \approx 6274$ MeV), weak decay production mechanisms are strictly kinematically forbidden. They are most naturally searched for in prompt production at high-energy hadron colliders, especially at LHCb, similar to the prompt production mechanisms of the $X(3872)$ and $T_{cc}^+$ states. In Table~\ref{tab:exp-search} the most relevant threshold regions, suggested final states, and expected line-shape features for the predicted poles are summarized. The table is understood as a qualitative guide for experimental searches rather than a prediction of absolute production rates, since the production mechanism is not included in the present OBE calculation.

\begin{table*}[htbp]
\centering
\caption{Suggested experimental search channels and characteristic signals for the predicted $P_{\bar b s}$ molecular candidates.}
\label{tab:exp-search}
\begin{ruledtabular}
\begin{tabular}{c c c c}
$J^P$(Dominant channel) & Threshold region(MeV) & Suggested final states & Expected signal \\
\hline
$1/2^-(B\Sigma)$ 
& $6472.8$ 
& $B_s^0N$, $B\Lambda$ 
& near-threshold peak \\
$1/2^-(B^*\Sigma)$ 
& $6517.9$ 
& $B\Lambda$, $B^*\Lambda$ 
& cusplike line shape \\

$3/2^-(B^*\Sigma)$ 
& $6517.9$ 
& $B^*\Lambda$ 
& near-threshold peak\\

$3/2^-(B\Sigma^*)$ 
& $6662.5$ 
& $B\Lambda,B^*\Lambda,B_s^0N$ 
& near-threshold peak \\

$1/2^-(B^*\Sigma^*)$ 
& $6707.6$ 
& $B\Lambda,B_s^0N$ 
& near-threshold peak \\
$3/2^-(B^*\Sigma^*)$ 
& $6707.6$ 
& $B\Lambda,B_s^0N$ 
& brought peak \\
\end{tabular}
\end{ruledtabular}
\end{table*}

\section{Summary}
In this work, we have studied the $B_s^0N- B^{(*)}\Lambda- B^{(*)}\Sigma$ coupled channel system with isospin $I=1/2$ using the OBE model. The scalar-, pseudoscalar-, and vector-meson-exchange potentials were derived from effective Lagrangians constrained by HQSS and SU(3) flavor symmetry. The coupled-channel Schr\"odinger equation was solved with S-D wave mixing, and the poles were extracted by analytically continuing the coupled-channel S matrix to the complex energy plane. To examine the uncertainty from unresolved short-range dynamics, we use a cutoff $\Lambda$ and a dimensionless parameter $a$, which controls the contribution of the short-range $\delta(\bm r)$ term.

The single-channel analysis shows that the $B_s^0N$, $ B\Lambda$, and $ B^*\Lambda$ interactions are not sufficiently attractive to form bound states in the considered parameter region. The $ B\Sigma$ channel with $J^P=1/2^-$ forms a bound state and is almost insensitive to the parameter $a$, indicating that its binding mechanism is mainly controlled by the long-range part of the OBE potential. By contrast, the $ B^*\Sigma$ channels with $J^P=1/2^-$ and $3/2^-$ show a pronounced dependence on the short-range interaction. The opposite behavior of these two spin channels under variations of $a$ reflects the different roles played by the contact term in the corresponding spin structures.

In the coupled-channel calculation, two poles in $J^P=1/2^-$ and a single pole in $J^P=3/2^-$ systems are found at $a=0$ in the RSs connected to physical real energy axis. In $1/2^-$ system, first pole appears below the $ B\Sigma$ channel threshold and couples most strongly to the $ B\Sigma$ channel, confirming its molecular origin from the $ B\Sigma$ interaction. Its dominant decay mode is the $B_s^0N$ channel, while the $ B\Lambda$ and $ B^*\Lambda$ channels provide complementary decay modes. The second pole in this system appears near the $ B^*\Sigma$ threshold. Its coupling is dominated by the $ B^*\Sigma(^{2}S_{1/2})$ component. This pole is strongly affected by short-range $\delta(r)$ term, it may appear experimentally as a cusplike structure rather than as a simple Breit-Wigner peak. The third pole appears in the $J^P=3/2^-$ system and is dominated by the $ B^*\Sigma(^{4}S_{3/2})$ component. The $B_s^0N$ decay channel mode is suppressed because it couples with the $D$ wave, whereas the $ B\Lambda$ and $ B^*\Lambda$ channels provide the dominant open decay contributions. This feature makes the $ B^*\Lambda$ and $ B\Lambda$ invariant-mass spectra especially relevant for testing this pole in $J^P=3/2^-$ quantum number. In addition, the effect of the higher $B\Sigma^*$ and $B^*\Sigma^*$ channels are studied, and which does not qualitatively change the three near-threshold poles associated with the original $B\Sigma$ and $B^*\Sigma$ thresholds. Additional $\Sigma^*$-related pole structures may appear near the $B\Sigma^*$ and $B^*\Sigma^*$ thresholds. Among them, the $J^P=3/2^-(B\Sigma^*)$ pole is relatively stable in the considered parameter region, whereas the $J^P=1/2^-(B^*\Sigma^*)$ and $J^P=3/2^-(B^*\Sigma^*)$ poles show stronger sensitivity to the short-range parameter $a$. 

Our results indicate that the three predicted poles lie in the $6.44-6.70$ GeV energy region and are relatively narrow. Their pole positions, pole couplings, and widths demonstrate that the spectrum is governed by the combined effects of near-threshold $ B^{(*)}\Sigma$ attraction, coupled-channel dynamics, S-D wave mixing, and short-range interactions. In prompt production at LHC, the amplitude analysis of $B_s^0N$ and $B^{(*)}\Lambda$ final states would provide the most direct test of the antibottom-strange molecular pentaquarks proposed in this work.

\begin{acknowledgments}	
We thank Yakefu Reyimuaji for useful discussion. This work is supported by the Natural Science Foundation of the Xinjiang Uyghur Autonomous Region of China under Grant No. 2025D01C292. The work of N. Y. is further supported by the National Natural Science Foundation of China under Grant No. 12565016.
\end{acknowledgments}

\section*{DATA AVAILABILITY}
    The experimental masses pf the particles that support the findings of this article are openly available at Review of Particle Physics in Ref.~\cite{ParticleDataGroup:2026aaa}.

\appendix

\section{ The potentials related with $ B\Sigma^*$ and $B^*\Sigma^*$ channels}\label{app:decuplet-pot}

The interaction vertices between decuplet bartons($T$) and the octet baryons($\mathcal{B})$ with light mesons are described by the following Lagrangian\cite{Matsuyama:2006rp,Ronchen:2012eg}  
\begin{align}
\mathcal{L}_{T\mathcal{B}\mathbb{P}}&=\frac{g_{T\mathcal{B}\mathbb{P}}}{m_\mathbb{P}}\bar T^\mu \partial_\mu \mathbb{P} \mathcal{B}+H.c.,\label{eq:lag-TBP}\\
\mathcal{L}_{T\mathcal{B}\mathbb{V}}&=i\frac{f_{T\mathcal{B}\mathbb{V}}}{m_\mathbb{V}}T^\mu \gamma^\nu \gamma_5 (\partial_\mu \mathbb{V}_\nu-\partial_\nu \mathbb{V}_\mu)\mathcal{B}+H.c.,\label{eq:lag-TBV}\\
\mathcal{L}_{TT\mathbb{P}}&=\frac{g_{TT\mathbb{P}}}{m_\mathbb{P}}\bar T_\alpha \gamma^\mu \gamma^5 \partial_\mu \mathbb{P} T^\alpha,\label{eq:lag-TTP}\\
\mathcal{L}_{TT\mathbb{V}}&=-g_{TT\mathbb{V}}\bar T_\alpha \gamma^\mu \mathbb{V}_\mu T^\alpha+\frac{f_{TT\mathbb{V}}}{2M_T}\bar T_\alpha \sigma^{\mu\nu}\partial_\nu \mathbb{V}_{\mu}T^\alpha,\label{eq:lag-TTV}\\
\mathcal{L}_{TT\sigma}&=g_{TT\sigma}\bar T_\alpha \sigma T^\alpha,\label{eq:lag-TTS}
\end{align}  
where $T^\mu$ is the decuplet baryon field with spin-$3/2$. In these Lagrangian, the flavor information is not explicitly included, and can be accounted by SU(3) singlet terms from the product representations of each Lagrangian with the SU(3) Clebsh-Gordan coefficients in Ref.~\cite{deSwart:1963pdg}. With prescription in Subsec.~\ref{sec:pot1}, the effective potentials for the transitions $\mathcal{\tilde{P}}^{(*)}T\to \mathcal{\tilde{P}}^{(*)}T$, $\mathcal{\tilde{P}}^{(*)}T\to \mathcal{\tilde{P}}^{(*)}\mathcal{B}$ are derived and listed as follows:
\begin{itemize}
	\item $\mathcal{\bar P}T\to \mathcal{\bar P}T $
	\begin{align}
		\mathcal{V}_4^\sigma&=-g_sg_{TT\sigma}\frac{\bm \chi_3^\dagger\cdot \bm \chi_1}{\bm q^2+m_\sigma^2},\\
		\mathcal{V}_4^\mathbb{V}&=-\frac{\sqrt2}{2}\beta g_Vg_{TT\mathbb{V}}\frac{\bm \chi_3^\dagger\cdot \bm \chi_1}{\bm q^2+m_\mathbb{V}^2},
	\end{align}
		\item $\mathcal{\bar P}^*\mathcal{B} \to\mathcal{\bar P}T$
	\begin{align}
		\mathcal{V}_5^\mathbb{P}&=\frac{gg_{T\mathcal{B}\mathbb{P}}}{f_\pi m_P}\frac{\bm \epsilon_2\cdot \bm q \bm \chi_3^\dagger \cdot \bm q}{\bm q^2+\mu_\mathbb{P}^2},\\
		\mathcal{V}_5^\mathbb{V}&=\frac{\sqrt2 \lambda g_vf_{T\mathcal{B}\mathbb{V}}}{M_\mathbb{V}}\notag\\
		&\times\frac{-\bm q^2(i\bm\chi_3^\dagger\times \bm \sigma)\cdot \bm\epsilon_2+(i\bm\chi_3^\dagger\times \bm \sigma)\cdot\bm q \bm\epsilon_2\cdot\bm q}{\bm q^2+\mu_\mathbb{V}^2},
	\end{align}
		\item $\mathcal{\bar P}\mathcal{B} \to\mathcal{\bar P}T$
	\begin{align}
		\mathcal{V}_6&=0,
	\end{align}	
		\item $\mathcal{\bar P}^*T\to \mathcal{\bar P}^*T$
	\begin{align}
		\mathcal{V}_7^\sigma&=g_sg_{TT\sigma}\frac{\bm\epsilon^\dagger_4\cdot \bm \epsilon_2 \bm \chi_3^\dagger \cdot \bm \chi_1}{\bm q^2+m_\sigma^2},\\
		\mathcal{V}_7^\mathbb{P}&=-\frac{gg_{TT\mathbb{P}}}{f_\pi m_\mathbb{P}}\frac{\bm q^2\bm \chi_3^{\dagger j}(i\bm \epsilon_4^\dagger\times \bm \epsilon_2)\cdot \bm \sigma\bm\chi_1^j}{\bm q^2+\mu_\mathbb{P}^2},\\
		\mathcal{V}_7^\mathbb{V}&=-\frac{\sqrt2 f_{TT\mathbb{V}} \lambda g_V }{2M_T}\frac{2\bm q^2\bm\chi_3^{\dagger j}(i\bm\epsilon_4^\dagger\times \bm\epsilon_2)\cdot \bm \sigma\bm \chi_1^j)}{\bm q^2+\mu_\mathbb{V}^2},
	\end{align}
		\item $\mathcal{\bar P}T\to\mathcal{\bar P}^*T$
	\begin{align}
		\mathcal{V}_8^\mathbb{P}&=\frac{gg_{TT\mathbb{P}}}{f_\pi m_\mathbb{P}}\frac{\bm q\cdot \bm\epsilon_4^\dagger \bm\chi_3^{\dagger j}\bm\sigma\cdot\bm q \bm\chi_1^j}{\bm q^2+\mu_\mathbb{P}^2},\\
		\mathcal{V}_8^\mathbb{V}&=\frac{f_{TT\mathbb{V}} \lambda g_V }{\sqrt2 M_T}\frac{\bm\chi_3^{\dagger j}(\bm\epsilon_4^\dagger\cdot\bm q \bm\sigma\cdot \bm q-\bm\epsilon_4^\dagger\cdot\bm\sigma\bm q^2)\bm\chi_1^j}{\bm q^2+\mu_\mathbb{V}^2},
	\end{align}
		\item $\mathcal{\bar P}^*\mathcal{B}\to\mathcal{\bar P}^*T$
	\begin{align}
		\mathcal{V}_9^\mathbb{P}&=-\frac{gg_{T\mathcal{B}P}}{f_\pi m_\mathbb{P}}\frac{\bm\chi_3^\dagger\cdot\bm q (i\bm\epsilon_4^\dagger\times\bm\epsilon_2)\cdot\bm q}{\bm q^2+\mu_\mathbb{P}^2},\\
		\mathcal{V}_9^\mathbb{V}&=\frac{2\sqrt2f_{T\mathcal{B}\mathbb{V}} \lambda g_V }{ M_\mathbb{V}}\frac{q^2(i\bm\chi_3^\dagger\times\bm\sigma)\cdot(i\bm\epsilon_4^\dagger\times\epsilon_2)}{\bm q^2+\mu_\mathbb{V}^2},
	\end{align}
		\item $\mathcal{\bar P}\mathcal{B}\to\mathcal{\bar P}^*T$
	\begin{align}
		\mathcal{V}_{10}^\mathbb{P}&=-\frac{gg_{T\mathcal{B}\mathbb{P}}}{f_\pi m_\mathbb{P}}\frac{\bm\chi_3^\dagger\cdot\bm q \bm\epsilon_4^\dagger\cdot\bm q}{\bm q^2+\mu_\mathbb{P}^2},\\
		\mathcal{V}_{10}^\mathbb{V}&=-\frac{\sqrt2f_{T\mathcal{B}\mathbb{V}} \lambda g_V }{ M_\mathbb{V}}\notag\\
		&\times\frac{\bm q^2(i\bm\chi_3^\dagger\times \bm \sigma)\cdot \bm\epsilon_4^\dagger-(i\bm\chi_3^\dagger\times \bm \sigma)\cdot\bm q \bm\epsilon_4^\dagger\cdot\bm q}{\bm q^2+\mu_\mathbb{V}^2}.
	\end{align}
\end{itemize}
With these potentials above, the potentials for $B_s^0N\to B^{(*)}\Sigma^*$, $B^{(*)}\Lambda\to B^{(*)}\Sigma^*$, $B^{(*)}\Sigma\to B^{(*)}\Sigma^*$ and $B^{(*)}\Sigma^*\to B^{(*)}\Sigma^*$ transitions are explicitly shown in Table~\ref{tab:poten-in-p-T}. 
\begin{table*}[ht!]\centering
	\caption{OBE potentials of $B^{(*)}\Sigma^*$ channels coupled to $B_s^0N- B^{(*)}\Lambda- B^{(*)}\Sigma$ system with $I=1/2$.}\label{tab:poten-in-p-T}  
	\begin{ruledtabular}
	\begin{tabular}{l c c c c c c c}
Transition                  & $\sum\limits_{\rm{ex}}\tau_{\rm{ex}}\mathcal{V}^{\rm{ex}}_i$                                                                                                            & $g_{T\mathcal{B}\mathbb{P}}$                & $f_{T\mathcal{B}\mathbb{V}}$            & $g_{TT\mathbb{P}}$                 & $g_{TT\mathbb{V}}$           & $f_{TT\mathbb{V}}$                & $g_{TT\sigma}$          \\
$B\Sigma^*\to B\Sigma^*$    & $\mathcal{V}_4^{\sigma}-\sqrt{2}\mathcal{V}_4^\rho+\frac{1}{\sqrt2}\mathcal{V}_4^\omega$                                                                                & 0                                           & 0                                       & 0                                  & $g_{\Delta\Delta\rho}$       & $f_{\Delta\Delta\rho}$            & $g_{\Sigma^*\Sigma^*\sigma}$ \\
$B^*\Sigma\to B\Sigma^*$    & $-\sqrt{2}\mathcal{V}_5^{\rho}+\frac{1}{\sqrt{2}}\mathcal{V}_5^\omega-\sqrt{2}\mathcal{V}_5^\pi+\frac{1}{\sqrt{2}}\mathcal{V}_5^\eta$                                   & $-\frac{1}{\sqrt6}g_{\Delta N\pi}$          & $-\frac{1}{\sqrt6}f_{\Delta N\rho}$     & 0                                  & 0                            & 0                                 & 0                     \\
$B^*\Lambda\to B\Sigma^*$   & $-\sqrt{\frac{3}{2}}\mathcal{V}_5^\rho-\sqrt{\frac{3}{2}}\mathcal{V}_5^\pi$                                                                                             & $-\frac{1}{\sqrt2}g_{\Delta N\pi}$          & $-\frac{1}{\sqrt2}f_{\Delta N\rho}$     & 0                                  & 0                            & 0                                 & 0                     \\
$B^*\Sigma^*\to B^*\Sigma^*$& $\mathcal{V}_7^{\sigma}-\sqrt2\mathcal{V}_7^{\rho}+\frac{1}{\sqrt2}\mathcal{V}_4^\omega-\sqrt{2}\mathcal{V}_7^\pi$                                                      & 0                                           & 0                                       & $g_{\Delta\Delta\pi}$              & $g_{\Delta\Delta\rho}$       & $f_{\Delta\Delta\rho}$            & $g_{\Sigma^*\Sigma^*\sigma}$ \\
$B\Sigma^*\to B^*\Sigma^*$  & $-\sqrt{2}\mathcal{V}_8^{\rho}+\frac{1}{\sqrt2}\mathcal{V}_4^\omega-\sqrt{2}\mathcal{V}_8^\pi$                                                                          & 0                                           & 0                                       & $g_{\Delta\Delta\pi}$              & $g_{\Delta\Delta\rho}$       & $f_{\Delta\Delta\rho}$            & 0 \\
$B^*\Sigma\to B^*\Sigma^*$  & $-\sqrt{2}\mathcal{V}_9^{\rho}+\frac{1}{\sqrt{2}}\mathcal{V}_9^\omega-\sqrt{2}\mathcal{V}_9^\pi+\frac{1}{\sqrt{2}}\mathcal{V}_9^\eta$                                   & $-\frac{1}{\sqrt6}g_{\Delta N\pi}$          & $-\frac{1}{\sqrt6}f_{\Delta N\rho}$     & $0$                                & 0                            & 0                                 & 0                     \\
$B^*\Lambda\to B^*\Sigma^*$ & $-\sqrt{\frac{3}{2}}\mathcal{V}_9^\rho-\sqrt{\frac{3}{2}}\mathcal{V}_9^\pi$                                                                                             & $-\frac{1}{\sqrt2}g_{\Delta N\pi}$          & $-\frac{1}{\sqrt2}f_{\Delta N\rho}$     & 0                                  & 0                            & 0                                 & 0                     \\
$B\Sigma\to B^*\Sigma^*$    & $-\sqrt{2}\mathcal{V}_{10}^{\rho}+\frac{1}{\sqrt{2}}\mathcal{V}_{10}^\omega-\sqrt{2}\mathcal{V}_{10}^\pi+\frac{1}{\sqrt{2}}\mathcal{V}_{10}^\eta$                       & $-\frac{1}{\sqrt6}g_{\Delta N\pi}$          & $-\frac{1}{\sqrt6}f_{\Delta N\rho}$     & 0                                  & 0                            & 0                                 & 0                     \\
$B\Lambda\to B^*\Sigma^*$   & $-\sqrt{\frac{3}{2}}\mathcal{V}_{10}^\rho-\sqrt{\frac{3}{2}}\mathcal{V}_{10}^\pi$                                                                                       & $-\frac{1}{\sqrt2}g_{\Delta N\pi}$          & $-\frac{1}{\sqrt2}f_{\Delta N\rho}$     & 0                                  & 0                            & 0                                 & 0                     \\
$B_s^0N\to B^*\Sigma^*$     & $-\sqrt{\frac{3}{2}}\mathcal{V}_{10}^{\bar{K}^*}$                                                                                                                       & 0                                           & $ \frac{1}{\sqrt3}f_{\Delta N\rho}$     & 0                                  & 0                            & 0                                 & 0                     \\
\end{tabular}
\end{ruledtabular}
\end{table*}
The relations of coupling constants are obtained by expanding the effective Lagrangian in Eqs.~\eqref{eq:lag-TBP}-\eqref{eq:lag-TTS} with the SU(3) Clebsh-Gordan coefficients in Ref.~\cite{deSwart:1963pdg}. The relations of relevant couplings with the $g_{\Delta N\pi}$ and $g_{\Delta\Delta \pi}$ are 
\begin{align}
	g_{\Sigma^*\Sigma\pi}&=\frac{1}{2\sqrt{30}}g_{T\mathcal{B}\mathbb{P}}=-\frac{1}{\sqrt{6}}g_{\Delta N\pi}, \notag\\
	g_{\Sigma^*\Lambda\pi}&=- \frac{1}{\sqrt{2}}g_{\Delta N\pi}, g_{\Sigma^* N \bar K}= \frac{1}{\sqrt{3}}g_{\Delta N\pi}, \notag\\
	g_{\Sigma^*\Sigma^*\pi}&=-\frac{1}{2\sqrt{15}}g_{TT\mathbb{P}}=g_{\Delta\Delta \pi}, 
\end{align}
where $g_{\Delta N\pi}=-\frac{1}{2\sqrt{5}}g_{T\mathcal{B}\mathbb{P}}=2.13$\cite{Ronchen:2012eg}, $g_{\Delta \Delta \pi}=-\frac{1}{2\sqrt{15}}g_{TT\mathbb{P}}=1.78$\cite{Schutz:1998jx}. The signs of these couplings are checked with the technics in the Appendix B of Ref.~\cite{Yalikun:2021dpk}, and are consistent with the quark model. These relations are explicitly shown with following Lagrangian: the singlet terms of the $10^*_T\times 8_P\times 8_{\mathcal{B}}$, $10^*_T\times 8_P\times 10_T$, and $10^*_T\times 10_T\times 1_\sigma$ product representations in the isospin bases are shown as below 
\begin{align}
	\frac{\mathcal{L}_{T\mathcal{B}\mathbb{P}}}{g_{T\mathcal{B}\mathbb{P}}}&\to -\frac{1}{2\sqrt5}\bar{\Delta}S_a^\dagger \pi_a N
	+ \frac{1}{2\sqrt{10}} \bar{\Sigma}^* t_{a} \pi_a\Lambda\notag\\
	&- \frac{1}{2\sqrt{15}} \bar{\Sigma}^* u_{b} K_a^c N+ \frac{1}{2\sqrt{30}}  \bar{\Sigma}^* (I_a\pi_a+\sqrt{3}\eta)\Sigma\notag\\
&+ \frac{1}{2\sqrt{30}} \bar{\Xi}^* (\tau_a \pi_a+\sqrt{3}\eta) \Xi + \frac{1}{2\sqrt{15}}\bar{\Delta}r_{b} K_a \Sigma \notag\\
&- \frac{1}{2\sqrt{10}} \bar{\Xi}^*  K^c \Lambda 
- \frac{1}{2\sqrt{15}} \bar{\Xi}^* u^T_{b}  K_a^\dagger \Sigma\notag\\
&
+ \frac{1}{2\sqrt{15}} \bar{\Sigma}^*u_{b}\bar K_a^\dagger  \Xi 
- \frac{1}{2\sqrt{5}}\bar{\Omega}K^\dagger \Xi  
+ H.c.,\\
\frac{\mathcal{L}_{TT\mathbb{P}}}{g_{TT\mathbb{P}}}
&\to -\frac{1}{2\sqrt{15}}\bar{\Delta}(Z_a \pi_a+\frac{\sqrt{3}}{2}\eta)\Delta \notag\\
&- \frac{1}{2\sqrt{15}}\Sigma^* I_a \pi_a \Sigma^*+\frac{1}{2\sqrt{30}}\bar\Delta r_{b}K_a \Sigma^*
\notag\\
&+\frac{1}{\sqrt{30}}\bar\Sigma^* u_{b} K_a \Xi^*
+\frac{1}{2\sqrt{10}}\bar\Xi^* K \Omega\notag\\
&+\frac{1}{2\sqrt{30}}\Sigma^*r_{b}^TK_a^\dagger \Delta
+\frac{1}{\sqrt{30}}\bar\Xi^* u_{b}^T  K_a^\dagger \Sigma^*
\notag\\
&-\frac{1}{4\sqrt{15}}\bar{\Xi}^*(\tau_a \pi_a-\frac{\sqrt{3}}{2}\eta)\Xi^* \notag\\
&+\frac{1}{2\sqrt{10}}\bar{\Omega} K^\dagger\Xi^*+\frac{1}{2\sqrt{5}}\bar{\Omega}\eta\Omega,\\
\frac{\mathcal{L}_{TT\sigma}}{g_{TT\sigma}}&\to -\frac{1}{\sqrt{10}}(\bar\Delta\sigma\Delta+\bar\Sigma^*\sigma\Sigma^*+\bar{\Xi}^*\sigma\Xi^*+\bar{\Omega}\sigma\Omega),
\end{align}
where the $S_a^\dagger$ is the spin $3/2\to 1/2$ transition matrix; $I_a$, $\tau_a$ and $Z_a$ are the standard spin $1$, $1/2$ and $3/2$ matrices, respectively. The matrix form of them reads
\begin{align}
I_1 &= \frac{1}{\sqrt{2}}
\begin{pmatrix}
0 & 1 & 0 \\
1 & 0 & 1 \\
0 & 1 & 0
\end{pmatrix}, I_2 = \frac{1}{\sqrt{2}}
\begin{pmatrix}
0 & -i & 0 \\
i & 0 & -i \\
0 & i & 0
\end{pmatrix},\notag\\
	I_3 &=
\begin{pmatrix}
1 & 0 & 0 \\
0 & 0 & 0 \\
0 & 0 & -1
\end{pmatrix},\notag\\
\tau_1 &= \frac{1}{2}
\begin{pmatrix}
0 & 1 \\
1 & 0
\end{pmatrix}, \tau_2 = \frac{1}{2}
\begin{pmatrix}
0 & -i \\
i & 0
\end{pmatrix}, \tau_3 = \frac{1}{2}
\begin{pmatrix}
1 & 0 \\
0 & -1
\end{pmatrix},\notag\\
Z_1 &= 
\begin{pmatrix}
0 & \frac{\sqrt{3}}{2} & 0 & 0 \\
\frac{\sqrt{3}}{2} & 0 & 1 & 0 \\
0 & 1 & 0 & \frac{\sqrt{3}}{2} \\
0 & 0 & \frac{\sqrt{3}}{2} & 0
\end{pmatrix}, \notag\\
Z_2 &= 
\begin{pmatrix}
0 & -\frac{\sqrt{3}i}{2} & 0 & 0 \\
\frac{\sqrt{3}i}{2} & 0 & -i & 0 \\
0 & i & 0 & -\frac{\sqrt{3}i}{2} \\
0 & 0 & \frac{\sqrt{3}i}{2} & 0
\end{pmatrix}, \notag\\
Z_3 &=
\begin{pmatrix}
\frac{3}{2} & 0 & 0 & 0 \\
0 & \frac{1}{2} & 0 & 0 \\
0 & 0 & -\frac{1}{2} & 0 \\
0 & 0 & 0 & -\frac{3}{2}
\end{pmatrix}, S_1^\dagger =\begin{pmatrix}
-\dfrac{1}{\sqrt{2}} & 0 \\
0 & -\dfrac{1}{\sqrt{6}} \\
\dfrac{1}{\sqrt{6}} & 0 \\
0 & \dfrac{1}{\sqrt{2}}
\end{pmatrix},\notag\\
S_2^\dagger &= \begin{pmatrix}
\dfrac{i}{\sqrt{2}} & 0 \\
0 & \dfrac{i}{\sqrt{6}} \\
\dfrac{i}{\sqrt{6}} & 0 \\
0 & \dfrac{i}{\sqrt{2}}
\end{pmatrix}, S_3^\dagger = \begin{pmatrix}
0 & 0 \\[2pt]
\sqrt{\dfrac{2}{3}} & 0 \\[2pt]
0 & \sqrt{\dfrac{2}{3}} \\[2pt]
0 & 0
\end{pmatrix}.\notag
\end{align}
For convenience of calculation isospin factor, we also define the  $t_a$, $u_b$ and $r_b$ matrices,
\begin{align}
	t_1&=\begin{pmatrix}
		\frac{1}{\sqrt2}\\
		0\\
		-\frac{1}{\sqrt2}
	\end{pmatrix}, 
	t_2=\begin{pmatrix}
		-\frac{i}{\sqrt2}\\
		0\\
		-\frac{i}{\sqrt2}
	\end{pmatrix},
	t_3=\begin{pmatrix}
		0\\
		-1\\
		0
	\end{pmatrix},\notag\\
u_1&=\begin{pmatrix}
		1&0\\
		0&\frac{1}{\sqrt2}\\
		0&0
	\end{pmatrix}, u_2=\begin{pmatrix}
		0&0\\
		\frac{1}{\sqrt2}&0\\
		0&1
	\end{pmatrix},\notag\\
	r_1&=\begin{pmatrix}
		\sqrt{3}&0&0\\
		0&\sqrt{2}&0\\
		0&0&1\\
		0&0&0
	\end{pmatrix}, 	r_2=\begin{pmatrix}
		0&0&0\\
		1&0&0\\
		0&\sqrt{2}&0\\
		0&0&\sqrt{3}
	\end{pmatrix},
\end{align}
where $t_a\pi_a=t_1\pi_1+t_2\pi_2+t_3\pi_3$, $u_bK_b=u_1K^++u_2K^0$, $r_b K_b=r_1K^++r_2K^0$.


\bibliographystyle{apsrev4-2}
\bibliography{article.bib}

\end{document}